\pdfoutput=1
\documentclass[11pt,twoside,a4paper,cmspaper,final,collab]{cms-tdr}
\def\svnVersion{35de786}\def\svnDate{2023/12/05}\def\cmsCernNoTag{CERN-EP-2023-105}\def\cmsCernDate{\today}\def\cmsMessage{Published in the European Physical Journal C as \href{http://dx.doi.org/10.1140/epjc/s10052-023-12258-4}{\doi{10.1140/epjc/s10052-023-12258-4}.}}
\begin{document}\cmsNoteHeader{SMP-21-005}

\newcommand{\MN}{\ensuremath{\PGm\PGn}}
\newcommand{\EN}{\Pe\PGn}
\newcommand{\IComb}{I_{\text{comb}}}
\newcommand{\pp}{\ensuremath{{\Pp\Pp}}}
\newcommand{\rts}{\ensuremath{\sqrt{s}}}
\providecommand{\MT}{\ensuremath{m_\text{T}}\xspace}
\newcommand{\Zll}{\ensuremath{\PZ \to \ell \ell}}
\newcommand{\Wen}{\ensuremath{\PW \to \EN}}
\newcommand{\Rcpm}{\ensuremath{R_\PQc^{\pm}}}
\newcommand{\Wmn}{\ensuremath{\PW \to \MN}}
\newcommand{\Wtn}{\ensuremath{\PW \to \PGt \PGn}}
\newcommand{\Wj}{\ensuremath{\PW+\text{jets}}}
\newcommand{\Wc}{\ensuremath{\PW+\PQc}}
\newcommand{\Wcc}{\ensuremath{\PW + \PQc \PAQc}}
\newcommand{\Wbb}{\ensuremath{\PW + \PQb  \PAQb }}
\newcommand{\Wb}{\ensuremath{\PW + \PQb}}
\newcommand{\Wlight}{\ensuremath{\PW + {\PQu\PQd\PQs\Pg}}\xspace}
\newcommand{\WQQ}{\ensuremath{\PW + \PQQ\PAQQ}\xspace}
\newcommand{\Zj}{\ensuremath{\PZ+\text{jets}}}
\newcommand{\PWmc}{\ensuremath{\PWm+\PQc}}
\newcommand{\PWpc}{\ensuremath{\PWp+\PAQc}}
\newcommand{\SWpc}{\sigma(\PWpc)}
\newcommand{\SWmc}{\sigma(\PWmc)}
\newcommand{\SWc}{\ensuremath{\sigma(\Wc)}}
\newcommand{\SWcdifflineeta}{\ensuremath{\rd\sigma(\Wc)/\rd\abs{\eta^\ell} }}
\newcommand{\SWcdifflinept}{\ensuremath{\rd\sigma(\Wc)/\rd{\pt^\ell}}}
\newcommand{\jet}{\text{jet}}
\newcommand{\cjet}{\PQc\text{ jet}\xspace}
\newcommand{\ppWc}{\ensuremath{\Pp\Pp \to \PW+\PQc}}
\newcommand{\ppWpc}{\ensuremath{\Pp\Pp \to \PWp+\PAQc}}  
\newcommand{\ppWmc}{\ensuremath{\Pp\Pp \to \PWm+\PQc}}  
\newcommand{\noppWc}{\ensuremath{\PW+\PQc}}
\newcommand{\OSSS}{\ensuremath{\text{OS-SS}}\xspace}
\DeclareRobustCommand{\PGLpmc}{{\HepParticle{\PGL}{c}{\pm}}\Xspace}

\newlength\cmsTabSkip\setlength{\cmsTabSkip}{1ex}
\newcommand{\PYTHIAeight}{\PYTHIA{8}\xspace}
\ifthenelse{\boolean{cms@external}}{\providecommand{\cmsLeft}{upper\xspace}}{\providecommand{\cmsLeft}{left\xspace}}
\ifthenelse{\boolean{cms@external}}{\providecommand{\cmsRight}{lower\xspace}}{\providecommand{\cmsRight}{right\xspace}}
\newlength\cmsFigWidth
\ifthenelse{\boolean{cms@external}}{\setlength{\cmsFigWidth}{0.49\textwidth}}{\setlength{\cmsFigWidth}{0.49\textwidth}}

\cmsNoteHeader{SMP-21-005}
\title{Measurement of the production cross section for a \texorpdfstring{\PW}{W} boson in association with a charm quark in proton-proton collisions at \texorpdfstring{$\rts=13\TeV$}{sqrt(s) = 13 TeV}}
\titlerunning{Measurement of the production of a \texorpdfstring{\PW}{W} boson in association with a charm quark at 13\TeV}
\date{\today}

\abstract{
The strange quark content of the proton is probed through the measurement of the production cross section for a \PW boson and a charm (\PQc) quark in proton-proton collisions at a center-of-mass energy of 13\TeV. The analysis uses a data sample corresponding to a total integrated luminosity of 138\fbinv collected with the CMS detector at the LHC. The W bosons are identified through their leptonic decays to an electron or a muon, and a neutrino. Charm jets are tagged using the presence of a muon or a secondary vertex inside the jet. The $\Wc$ production cross section and  the  cross  section  ratio $\Rcpm = \SWpc/\SWmc$ are measured inclusively and differentially as functions of the transverse momentum and the pseudorapidity of the lepton originating from the \PW boson decay. The precision of the measurements is improved with respect to previous studies, reaching 1\% in $\Rcpm = 0.950 \pm 0.005\stat \pm 0.010 \syst$. The measurements are compared with theoretical predictions up to next-to-next-to-leading order in perturbative quantum chromodynamics.
}

\hypersetup{
pdfauthor={CMS Collaboration},
pdftitle={Measurement of the associated production of a W boson and a charm quark at sqrt(s)= 13 TeV},
pdfsubject={CMS},
pdfkeywords={CMS, SMP, W+charm}} 

\maketitle

\section{Introduction~\label{sec:intro}}
\begin{figure*}
  \centering
  \includegraphics[width=0.99\textwidth]{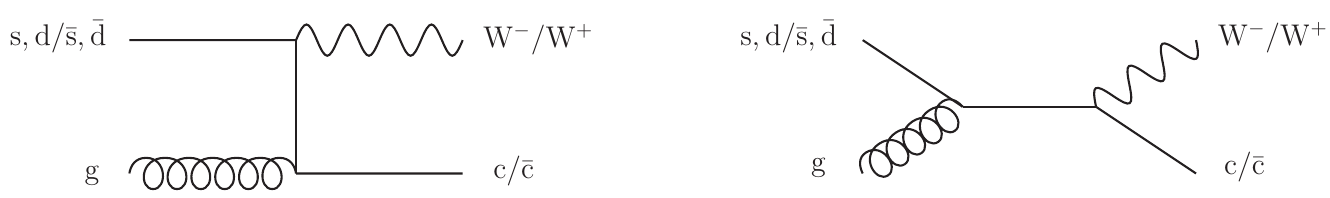}
  \caption{Leading order Feynman diagrams for the associated production of a \PW boson and a charm quark. The electric charges of the \PW boson and \PQc quark have opposite signs.}
  \label{fig:OSSS_diagram}
\end{figure*}

The associated production of a \PW boson and a single charm (\PQc) quark ($\Wc$) in proton-proton (\pp) collisions at the CERN LHC is directly sensitive to the strange quark ($\PQs$) content of the colliding protons at an energy scale of the order of the \PW  boson mass~\cite{Baur}. This sensitivity comes from the dominance of the $\PQs\Pg \to \Wc$ contribution over the Cabibbo-suppressed process $\PQd\Pg \to \Wc$ at tree level (see Fig.~\ref{fig:OSSS_diagram}). Therefore, this process provides valuable information on the strange quark parton distribution function~(PDF), which is one of the least constrained PDFs of the proton. Accurate measurements of the $\Wc$ production cross section and of the $\Rcpm = \SWpc/\SWmc$ cross section ratio can be used 
to further constrain the strange quark PDF,
and to probe the level of asymmetry between the $\PQs$ and $\PAQs$ PDFs~\cite{strange-pasym,strange-asym,strangeProton}.

Furthermore, the production of $\Wc$ events provides a useful calibration sample for the measurements and searches at the LHC involving electroweak bosons and \PQc quarks in the final state~\cite{EPJC78, CMS-PAPER-HIG-21-008}. 
Precise measurements of $\Wc$ production can be used to check the theoretical calculations of this process and its modeling in the currently available Monte Carlo (MC) event generators.

The $\Wc$ production in $\Pp\Pp$ collisions at the LHC has been reported by the CMS~\cite{CMS-PAPER-SMP-12-002, CMS-PAPER-SMP-18-013, CMS-PAPER-SMP-17-014}, ATLAS~\cite{WplusCATLAS,ATLAS-wc-13TeV-2023}, and LHCb~\cite{LHCb-HF} Collaborations at center-of-mass energies $\rts = 7$, 8, and 13\TeV. Measurements of $\Wc$ fiducial cross sections and the $\Rcpm$ cross section ratio were performed in those analyses by identifying charm events through the reconstruction of exclusive decays of charm hadrons, or finding secondary vertices  or muons inside a jet.

In this paper, we present a measurement of the $\Wc$ production cross section and cross section ratio $\Rcpm$ at  $\rts=13\TeV$ using the data collected in 2016--2018. The precision is improved compared with previous CMS measurements. 
In particular, the uncertainty in the $\Rcpm$ measurement is halved, reaching a precision of 1\%. 
Measurements are performed in four independent channels, 
depending on the method used for identifying the \PQc quarks and the \PW boson decay mode (electron or muon).
Jets are tagged as originating from the hadronization of \PQc quarks (\cjet) by the presence of either muons or secondary vertices inside the jets. The combination of the measurements in the four channels, the use of the large data set collected at $\rts=13\TeV$, and the reduction of systematic uncertainties, lead to more precise measurements.

A key property of $\Wc$ production is the opposite sign of the electric charges of the \PW boson and \PQc quark. This feature allows the suppression of most of the background events, which exhibit bottom or charm quarks and antiquarks with equal probability and identical kinematics, such as top quark-antiquark or $\PW+\PQc\PAQc$ production. The statistical subtraction of the distributions of physical observables for events where the reconstructed charges of the \PW boson and the \PQc quark have opposite sign (OS) and same sign (SS) leads to the effective removal of these backgrounds~\cite{CMS-PAPER-SMP-12-002, CMS-PAPER-SMP-18-013}.
This technique, referred to as \OSSS subtraction, enhances the sensitivity to the $\PQs\Pg \to \Wc$ process, and therefore to the strange quark PDF.

The \OSSS cross sections $\SWpc \equiv \sigma(\ppWpc){\mathcal{B}}(\PWp\to \ell^+  \PGn)$, $\SWmc \equiv \sigma(\ppWmc){\mathcal{B}}(\PWm\to \ell^-\PAGn)$ (where ${\mathcal{B}}$ denotes the branching fraction), their sum $\SWc \equiv \SWpc+\SWmc$, and the cross section ratio $\Rcpm \equiv \SWpc/\SWmc$ are measured. Inclusive and differential cross sections are measured as functions of the transverse momentum ($\pt^\ell$) and pseudorapidity ($\eta^\ell$) of the lepton from the \PW boson decay.  
Measurements are unfolded to the particle  and parton levels both in a fiducial region of phase space defined in terms of the kinematics of the lepton from the \PW boson ($\pt^{\ell} > 35\GeV$, $\abs{\eta^\ell} < 2.4$) and of the \cjet ($\pt^{\cjet}>30\GeV$, $\abs{\eta^{\cjet}}<2.4$).

The theoretical cross section for $\Wc$ production at the LHC~\cite{Stirling:2012vh} is well known at the next-to-leading order (NLO) accuracy in perturbative quantum chromodynamics (QCD). Recently, the first computation of next-to-NLO (NNLO) QCD corrections was published~\cite{NNLOWc,NNLOWc-ext}. The measurements presented here are compared with the predictions of these NNLO QCD calculations, which include NLO electroweak (EW) corrections.  
The measurements are also compared with the predictions of the parton-level MC program $\MCFM$~\cite{MCFM9}, which implements calculations at NLO in QCD using several proton PDF sets. 

The paper is structured as follows: the CMS detector is briefly described in Section~\ref{sec:CMS_det}, and the data and simulated samples used are
presented in Section~\ref{sec:samples}. Sections~\ref{sec:object_rec} and~\ref{sec:event_sel} describe the physics object reconstruction and the selection of the $\Wc$ signal sample. Section~\ref{sec:syst_uncert} reviews the most important sources of systematic uncertainties and their impact on the measurements. 
Cross section and cross section ratio measurements, compared with the NLO QCD theoretical predictions using different PDF sets, are detailed in Section~\ref{sec:xsec_meas}. The comparisons of the measurements with the NNLO QCD calculations are presented in Section~\ref{sec:xsec_NNLO}.
The main results of the paper are summarized in Section~\ref{sec:summary}. 

Tabulated results are provided in HEPData~\cite{HEPData}.

\section{The CMS detector~\label{sec:CMS_det}}
The central feature of the CMS apparatus is a superconducting solenoid
of 6\unit{m} internal diameter, providing a magnetic field of 3.8\unit{T}.
Within the magnetic volume are a silicon pixel
and strip tracker, a lead tungstate crystal electromagnetic calorimeter (ECAL),
and a brass and scintillator hadron calorimeter (HCAL),
each composed of a barrel and two endcap sections.
Additional forward calorimetry complements the coverage
provided by the barrel and endcap detectors.
The silicon tracker measures charged particles within the pseudorapidity
range $\abs{\eta}< 2.5$.
For nonisolated particles of $1 < \pt < 10\GeV$ and $\abs{\eta} < 1.4$, the
track resolutions are typically 1.5\% in $\pt$ and 20--75\mum
in the transverse impact parameter~\cite{CMS-DP-2020-032}. The upgrade of the pixel tracking detector~\cite{CMSpixel_upgrade} in early 2017, which includes additional layers and places the innermost layer closer to the interaction point, significantly improves the performance of heavy-flavor jet identification~\cite{Btagging13TeV}.
Muons are measured in the pseudorapidity range $\abs{\eta} < 2.4$, with detection planes made using three technologies: drift tubes, cathode strip chambers, and resistive plate chambers. 
A more detailed description of the CMS detector, together with a definition of the coordinate system used and the relevant kinematic variables, is reported in Ref.~\cite{CMS:2008xjf}.

Events of interest are selected using a two-tiered trigger system. The first level, composed of custom hardware processors, uses information from the calorimeters and muon detectors to select events at a rate of around 100\unit{kHz} within a fixed latency of about 4\mus~\cite{CMS:2020cmk}. The second level, known as the high-level trigger, consists of a farm of processors running a version of the full event reconstruction software optimized for fast processing, and reduces the event rate to around 1\unit{kHz} before data storage~\cite{CMS:2016ngn}.

\section{Data and simulated samples~\label{sec:samples}}

This analysis is performed using a data sample of $\pp$ collisions at $\rts=13\TeV$ collected by the CMS experiment during the 2016 ($36.3\fbinv$), 2017 ($41.5\fbinv$), and 2018 ($59.8\fbinv$) data-taking periods with a total integrated luminosity of $138\fbinv$.

The experimental signature of the signal events, an isolated high-$\pt$ lepton together with a \cjet,  is also present in other background processes. Sources of background include top quark production ($\ttbar$ and single top quark), diboson ($\PW\PW$, $\PW\PZ$, and $\PZ\PZ$) processes (collectively denoted as $\PV\PV$), the production of a \PZ boson (or a virtual photon) in association with jets ($\Zj$), and $\Wcc$ or $\Wbb$ events.

Samples of signal and background events are simulated using MC event generators based on fixed-order perturbative QCD calculations, supplemented with parton showering and multiparton interactions.
Simulated samples of $\Wj$ and $\Zj$ events are produced at NLO accuracy with the \MGvATNLO~\cite{Alwall} (version 2.6.3) matrix element generator with up to two partons in the final state. The decay of the \PW and \PZ bosons to tau leptons is included in the $\Wj$ and $\Zj$ simulations. 
Samples of $\ttbar$ and single top ($s$-, $t$-, and $\PQt\PW$ channels) events are generated at NLO accuracy with \POWHEG v2.0~\cite{Alioli:2010xd}. The cross sections for $\Wj$, $\Zj$, $\ttbar$, and single top production are obtained at NNLO in QCD~\cite{Li:2012wna,Czakon:2011xx}.  
The diboson production is modeled with samples of events generated with \PYTHIAeight~\cite{Sjostrand:2014zea} (version 8.219).

The simulated $\Wj$ sample is composed of \PW bosons accompanied by jets originating from quarks of all flavors and gluons.
Simulated $\Wj$ events are classified according to the flavor of the outgoing generated partons as: 
i) $\Wb$ if at least one bottom quark was generated in the hard process; ii) $\Wc$ if a single charm quark was created in the hard process; iii) $\Wcc$ if a $\PQc\PAQc$ pair was present in the event; iv) $\Wlight$ if no \PQc or \PQb quarks were produced.

{\tolerance=9000 
Data collected in different running periods are modeled with specific simulation configurations.
For simulations corresponding to 2016 detector conditions, the NLO NNPDF3.0~\cite{Ball:2014uwa} PDF set is used,  whereas the MC samples for 2017--2018 make use of the NNLO NNPDF3.1~\cite{NNPDF31nlo} PDF set.
The parton showering, hadronization, and the underlying events are modeled by \PYTHIA v8.212 (v8.230) using the CUETP8M1~\cite{Skands_2014,Khachatryan:2015pea} (CP5~\cite{Sirunyan_2020}) tune for the 2016 (2017--2018) samples.
The jet matching and merging scheme for the \MGvATNLO samples is FxFx~\cite{Frixione}.
\par}

In the \PYTHIAeight simulations, the charm fragmentation fractions, defined as the probabilities for \PQc quarks to hadronize 
as particular charm hadrons, corresponding to $\PDpm$, $\PDz/\PADz$, \PDpms and \PGLpmc hadrons, are corrected to match those in Ref.~\cite{Lisovyi:2015uqa}. In addition, the leptonic and hadronic decay branching fractions of those hadrons are corrected to agree with more recent measurements~\cite{PDG22}.  

{\tolerance=9000 
Generated events are processed through a full \GEANTfour-based~\cite{Agostinelli:2002hh} CMS detector simulation and trigger emulation.
Simulated events are reconstructed with the same algorithms used to reconstruct collision data.
\par}

{\tolerance=800 
The simulated samples incorporate additional $\pp$ interactions in the same or nearby bunch crossings (pileup) to reproduce the experimental conditions. Simulated events are weighted so the pileup distribution
matches the experimental data. 
\par}

\section{Object reconstruction~\label{sec:object_rec}}

The global event reconstruction (also called particle-flow event reconstruction~\cite{CMS:2017yfk}) reconstructs and identifies each individual particle in an event, with an optimized combination of all subdetector information. In this process, the identification of the particle type (photon, electron, muon, charged or neutral hadron) plays an important role in the determination of the particle direction and energy. Photons are identified as ECAL energy clusters not linked to the extrapolation of any charged-particle trajectory. Electrons are identified as a primary charged-particle track and with many ECAL energy clusters corresponding to this track extrapolation to the ECAL and to possible bremsstrahlung photons emitted along the path through the tracker material. Muons are identified as tracks in the central tracker consistent with either a track or several hits in the muon system, and associated with calorimeter deposits compatible with the muon hypothesis. Charged hadrons are identified as charged particle tracks neither identified as electrons nor as muons. Finally, neutral hadrons are identified as HCAL energy clusters not linked to any charged-hadron trajectory, or as a combined ECAL and HCAL energy excess with respect to the expected charged-hadron energy deposit.

The primary vertex (PV) is taken to be the vertex corresponding to the hardest scattering in the event, evaluated using tracking information alone, as described in Section 9.4.1 of Ref.~\cite{CMS-TDR-15-02}.

{\tolerance 800
The energy of photons is obtained from the ECAL measurement. The energy of electrons is determined from a combination of the track momentum at the PV, the corresponding ECAL cluster energy, and the energy sum of all bremsstrahlung photons associated with the track. The energy of muons is obtained from the corresponding track momentum. The energy of charged hadrons is determined from a combination of the track momentum and the corresponding ECAL and HCAL energies, corrected for the response function of the calorimeters to hadronic showers. Finally, the energy of neutral hadrons is obtained from the corresponding corrected ECAL and HCAL energies.
\par}

The electron momentum is estimated by combining the energy measurement in the ECAL with the momentum measurement in the tracker. The momentum resolution for electrons with $\pt \approx 45\GeV$ from $\PZ \to \Pe \Pe$ decays ranges from 1.6 to 5.0\%. It is generally better in the barrel region than in the endcaps, and also depends on the bremsstrahlung energy emitted by the electron as it traverses the material in front of the ECAL~\cite{CMS:2020uim,CMS-DP-2020-021}.
Matching muons to tracks measured in the silicon tracker results in a $\pt$ resolution of 1\% in the barrel and 3\% in the endcaps for muons with \pt up to 100\GeV~\cite{CMS:2018rym}. 

For each event, hadronic jets are clustered from these reconstructed particles using the infrared- and collinear-safe anti-\kt algorithm~\cite{Cacciari:2008gp, Cacciari:2011ma} with a distance parameter of 0.4. The jet momentum is determined as the vector sum of all particle momenta in the jet, and is found from simulation to be, on average, within 5--10\% of the true momentum over the entire \pt spectrum and detector acceptance. Pileup interactions can contribute additional tracks and calorimetric energy depositions to the jet momentum. To mitigate this effect, charged particles identified as originating from pileup vertices are discarded, and an offset correction is applied to correct for remaining contributions from neutral particles. Jet energy corrections are derived from simulation to bring the measured response of jets to that of particle level jets on average. In situ measurements of the momentum balance in dijet, $\text{photon} + \text{jet}$, $\PZ + \text{jet}$, and multijet events are used to account for any residual differences in the jet energy scale (JES) between data and simulation~\cite{CMS:2016lmd}. The jet energy resolution (JER) amounts typically to 15--20\% at 30\GeV, 10\% at 100\GeV, and 5\% at 1\TeV. Additional selection criteria are applied to each jet to remove jets potentially dominated by anomalous contributions from various subdetector components or reconstruction failures.

The missing transverse momentum vector $\ptvecmiss$ is the projection on the plane perpendicular to the beams of the negative vector momenta sum of all particles that are reconstructed with the particle-flow algorithm. The $\ptvecmiss$ is modified to account for corrections to the energy scale of the reconstructed jets in the event. The missing transverse momentum, $\ptmiss$, is defined as the magnitude of the $\ptvecmiss$
vector, and it is a measure of the transverse momentum of particles leaving the detector undetected~\cite{CMS:2019ctu}.

The trigger, reconstruction, and selection efficiencies are corrected in simulations to match those observed in the data.
Lepton efficiencies ($\epsilon_{\ell}$) are evaluated with data samples of dilepton events in the $\PZ$ boson mass peak with the tag-and-probe method~\cite{CMS-PAPER-EWK-10-005}, and
correction factors $\epsilon_{\ell}^\text{data}/\epsilon_{\ell}^{\mathrm{MC}}$, binned in $\pt^\ell$ and $\eta^\ell$ of the leptons, are implemented.

\section{Event selection~\label{sec:event_sel}}

Events with a high-$\pt$ lepton from the decay of a {\PW} boson are selected online by a trigger algorithm that  requires the presence of an electron (muon) candidate with minimum $\pt$ of 27, 32, and 32 $\GeVns$ (24, 27, and 24 $\GeVns$)
during the 2016, 2017, and 2018 data-taking periods, respectively.  Electrons and muons are selected using tight identification criteria following the reconstruction algorithms discussed in Refs.~\cite{CMS:2020uim,CMS:2018rym}.
The analysis follows the selection strategy used in Ref.~\cite{CMS-PAPER-SMP-18-013} and requires the presence of a high-$\pt$ isolated lepton in the region $\abs{\eta^{\ell}} < 2.4$ and $\pt^{\ell} > 35\GeV$. 

{\tolerance=800 
The combined isolation variable, $\IComb$, quantifies additional hadronic activity around the selected leptons.
It is defined as the sum of the transverse momenta of neutral hadrons, photons, and charged hadrons in a cone with
$\Delta R = \sqrt{\smash[b]{(\Delta\eta)^2 +(\Delta\phi)^2}}<0.3$ (0.4) around the electron (muon) candidate, excluding the contribution from the lepton itself, where $\phi$ is the azimuthal angle in radians.
Only charged particles originating from the PV are included in the sum to minimize the contribution from pileup interactions.
The contribution of neutral particles from pileup vertices is estimated and subtracted from $\IComb$.
For electrons, this contribution is evaluated with the jet area method described in Ref.~\cite{jet_area}; for muons,
it is assumed to be half the $\pt$ sum of all charged particles in the cone originating from pileup vertices.
The factor one-half accounts for the expected ratio of neutral to charged particle production in hadronic interactions.
The lepton candidate is considered to be isolated if $\IComb/\pt^{\ell} < 0.15$.
Events with an additional isolated lepton with $\pt^\ell>20\GeV$ are rejected  
to suppress the contribution from $\Zj$ and $\ttbar$ events. 
\par}

The transverse mass ($\MT$) of the lepton and $\ptvecmiss$ is defined as,
\begin{linenomath*}
\begin{equation*}
  \MT \equiv \sqrt{2~\pt^\ell~\ptmiss~[1-\cos(\phi_\ell-\phi_{\ptmiss})]},
\end{equation*}
\end{linenomath*}
where $\phi_\ell$ and $\phi_{\ptmiss}$ are the azimuthal angles of the lepton and the $\ptvecmiss$ vector. 
Events with $\MT < 55\GeV$ are discarded from the analysis to suppress the contamination from events composed uniquely of jets produced through the strong interaction, referred to as QCD multijet events. The contribution of this background was evaluated with two methods: (i) using a QCD multijet simulation; and (ii) by means of data control regions, inverting the selection requirements in transverse mass and lepton isolation to infer the contribution in the signal region. The contamination after \OSSS subtraction is negligible.

In addition to the requirements that select events with a \PW boson, we require the presence of at least one jet with $\pt^{\jet}>30\GeV$ and $\abs{\eta^{\jet}}<2.4$.
Jets with an angular separation between the jet axis and the selected isolated lepton $\Delta R ({\text{jet}},\ell)<0.4$ are not considered.

\subsection{Identification of charm jets~\label{sec:Wjetssel}}

Hadrons with \PQb and \PQc quark content decay through the weak interaction with lifetimes of the order of $10^{-12}\unit{s}$ and mean decay lengths larger than 100\mum
at the energies relevant for this analysis.
Secondary vertices well separated from the PV can be identified and reconstructed from the charged particle tracks.
In a sizeable fraction of the heavy-flavor hadron decays (${\approx}$10--15\%~\cite{PDG22}) there is a muon in the final state. 
We make use of these properties to define two independent data samples enriched with jets originating from a \PQc quark: i) the semileptonic (SL) channel, where a muon coming from the semileptonic decay of a \PQc hadron is identified inside a jet; and ii) the secondary vertex (SV) channel, where a displaced SV is reconstructed inside a jet.  
The charge of the \PQc quark is determined from the charge of the muon in the SL channel, and the charges of the SV tracks in the SV case, as described in more detail below.

If an event fulfills both the SL and SV selection requirements (about 6\% of the selected events), it is assigned to the SL channel. 
Thus, the SL and SV channels are mutually exclusive, \ie, the samples selected in each channel are statistically independent. 

These two signatures also feature weakly decaying \PQb hadrons.
Events from processes involving the associated production of \PW bosons and \PQb quarks
are abundantly selected in the two categories. The dominant background contribution stems from $\ttbar$ production, 
where a pair of \PW bosons and two \PQb jets are produced in the decays of the top quark-antiquark pair. 
This final state mimics the analysis topology when at least one of the \PW bosons decays leptonically and one of the \PQb jets contains an identified muon or a reconstructed SV.
However, this background is effectively suppressed by the \OSSS subtraction. A $\ttbar$ event will be categorized as OS (SS) when the lepton from the \PW decay and the muon or SV from the \PQb quark are coming from the same (different) top quark. 
The probability of identifying a muon or an SV inside the \PQb (or \PAQb) jet with opposite or same charge as the charge 
of the \PW candidate is expected to be the same, thus producing an equal amount of OS and SS events.

Top quark-antiquark events where one of the \PW bosons decays hadronically
into a $\PQc\PAQs$ (or $\PAQc\PQs$) pair may result in additional event candidates if the SL or SV signature originates from the \PQc jet. 
This topology produces genuine OS events, which contribute to the remaining background contamination after \OSSS subtraction. 
Similarly, single top quark production also produces OS events, but at a lower level because of the smaller production cross section. These remaining background contributions after \OSSS subtraction are estimated with simulations and are subtracted in the cross section measurements. 

The production of a \PW boson and a single bottom quark through the process $\Pq\Pg \to \Wb$, similar to the one sketched in Fig.~\ref{fig:OSSS_diagram},  
produces OS events, but it is heavily Cabibbo-suppressed and its contribution is negligible. The other source of a \PW boson and a \PQb quark is $\Wbb$ events 
where the $\bbbar$ pair originates from a gluon splitting mechanism. These events are also charge-symmetric, since it is equally likely to
identify the \PQb jet with the same or opposite charge than that of the \PW boson. This contribution also cancels out after the \OSSS subtraction. The same argument applies to $\Wcc$  events. 

\subsubsection{Event selection in the SL channel~\label{sec:Wsel_SL}}

The $\Wc$ events with a semileptonic \PQc quark decay are selected by requiring a reconstructed muon among the constituents of any of the selected jets.
Semileptonic \PQc quark decays into electrons are not considered because of a high background in identifying electrons inside jets.
The muon candidate must satisfy the same reconstruction and identification quality criteria as those imposed on the muons from the \PW boson decay, except for isolation, and must be reconstructed in the region 
$\abs{\eta} < 2.4$, with $\pt^{\PGm}<25\GeV$ and $\pt^{\PGm}/\pt^{\jet}<0.6$. The $\pt$ requirements reduce the contamination from prompt muons overlapping with or misreconstructed as jets.   
No minimum $\pt$ threshold is explicitly required, but the muon reconstruction algorithm sets a natural threshold of around 3 (2)\GeV in the barrel (endcap) region since the muon must traverse the material in front of the muon detector and penetrate deep enough into the muon system to be reconstructed and satisfy the identification criteria.
If more than one such muon is identified, the one with the highest $\pt$ is selected.

Additional requirements are applied for the event selection in the $\Wmn$ channel, since the selected sample is affected by a sizeable contamination from dimuon $\Zj$ events, where one of the muons from the $\PZ$ decay is reconstructed inside a jet. 
The track of the muon coming from a semileptonic decay of a charm hadron  tends to have a considerable transverse impact parameter with respect to the PV.
We require the transverse impact parameter significance (IPS) of the muon in the jet, defined as the muon transverse impact parameter divided by its uncertainty, to be larger than 2.
 In addition, events with a dimuon invariant mass
close to the $\PZ$ boson mass peak ($70<m_{\PGm\PGm}<110\GeV$) are discarded.
Furthermore, $m_{\PGm\PGm}$ must be larger than $12\GeV$ to suppress the background from low-mass resonances.

The normalizations of the $\ttbar$ and $\Zj$  backgrounds are derived from data control samples. A $\Zj$ data control sample is defined using the same selection criteria as the analysis but inverting the $m_{\PGm\PGm}$ requirement to select events close to the $\PZ$ boson mass peak ($70<m_{\PGm\PGm}<110\GeV$).
A normalization factor of $1.08\pm0.01$ is required to match the $\Zj$ simulation with data. The $\ttbar$ data control sample is established by selecting events with the same requirements as the analysis and additionally demanding at least three high-$\pt$ jets, two of which are tagged as $\cPqb$ jets (using the loose working point of the \textsc{DeepCSV} b-tagging algorithm~\cite{CMS-PAPER-BTV-16-002}), and the remaining jet contains a muon. A normalization factor of $0.92\pm0.02$ is required to bring into agreement data and $\ttbar$ simulation. The uncertainty in the background normalization factors reflects the statistical uncertainty of the data and the simulations in the control samples. Once the absolute normalization of the $\Zj$ and $\ttbar$ background contributions are determined, the $\Wj$ simulation is scaled so that the sum of the events from all predicted contributions be equal to the number of events in the selected data sample. The normalization factor of the $\Wj$ simulation (0.95) has only a minor effect in the contribution of the (small) predicted $\Wlight$ background. The overall normalization of the $\Wc$ signal simulation is irrelevant for the analysis, since it is only used for acceptance and efficiency calculations.

Events are classified as OS or SS depending on the electric charges of the lepton from the \PW boson decay and the muon inside the jet.

Table~\ref{yieldsSL} shows the event yields in the $\Wen$ and $\Wmn$ channels after the selection requirements described above and after \OSSS subtraction. The $\Wmn$ channel has a significantly lower yield due to the additional requirements to reduce the sizeable Z+jets background. The background yields, as estimated with the simulations, are also included in the table. The signal and background composition of the selected sample according to simulation is shown in Table~\ref{fractiontable0}.   
The fraction of signal $\Wc$ events in the $\Wen$ channel is above 80\%,
whereas in the $\Wmn$ channel it drops to  74\% because of the additional $\Zj$ background (around 6\%). 
The dominant background, $\ttbar$ production, where one of the \PW bosons 
decays leptonically and the other hadronically with a charm quark in the final state, amounts to approximately 10\%.

\begin{table}[htbp]
 \centering
  \topcaption{Data and background event yields (with statistical uncertainties) after selection and \OSSS subtraction for the SL channels (electron and muon \PW decay modes).}
  \label{yieldsSL}  \renewcommand{\arraystretch}{1.2}
  \begin{tabular}{ccc}
    SL channel  & Data  & Background  \\
    \hline
    $\Wen$ & 424\,047 $\pm$ 1286 & 80\,646 $\pm$ 933  \\
    $\Wmn$ & 263\,669 $\pm$ \,\,918  & 68\,108 $\pm$ 917 \\
  \end{tabular}
\end{table}

\begin{table*}[htbp]
 \centering
  \topcaption{Simulated signal and background composition (in percentage) of the SL sample after selection and \OSSS subtraction. The $\WQQ$ stands for the sum of the contributions of \Wcc~and \Wbb.}
  \label{fractiontable0}  \renewcommand{\arraystretch}{1.2}
  \begin{tabular}{lccccccc}
    SL channel     & \Wc  & \WQQ  & \Wlight  & $\Zj$  & \ttbar & Single \PQt  & $\PV\PV$  \\
    \hline
    $\Wen$ & 81.0 $\pm$ 0.6 & 0.5 $\pm$ 0.3 & 3.1 $\pm$ 0.5 & 0.4 $\pm$ 0.1 & 10.0 $\pm$ 0.1 & 4.4 $\pm$ 0.1 & 0.6 $\pm$ 0.1 \\
    $\Wmn$ & 74.2 $\pm$ 0.5 & 0.5 $\pm$ 0.3 & 2.0 $\pm$ 0.4 &  5.5 $\pm$ 0.2 &11.6 $\pm$ 0.1 & 5.8 $\pm$ 0.1 & 0.4 $\pm$ 0.1 \\
  \end{tabular}
\end{table*}

\begin{figure}[!tb]
\centering
\includegraphics[width=0.49\textwidth]{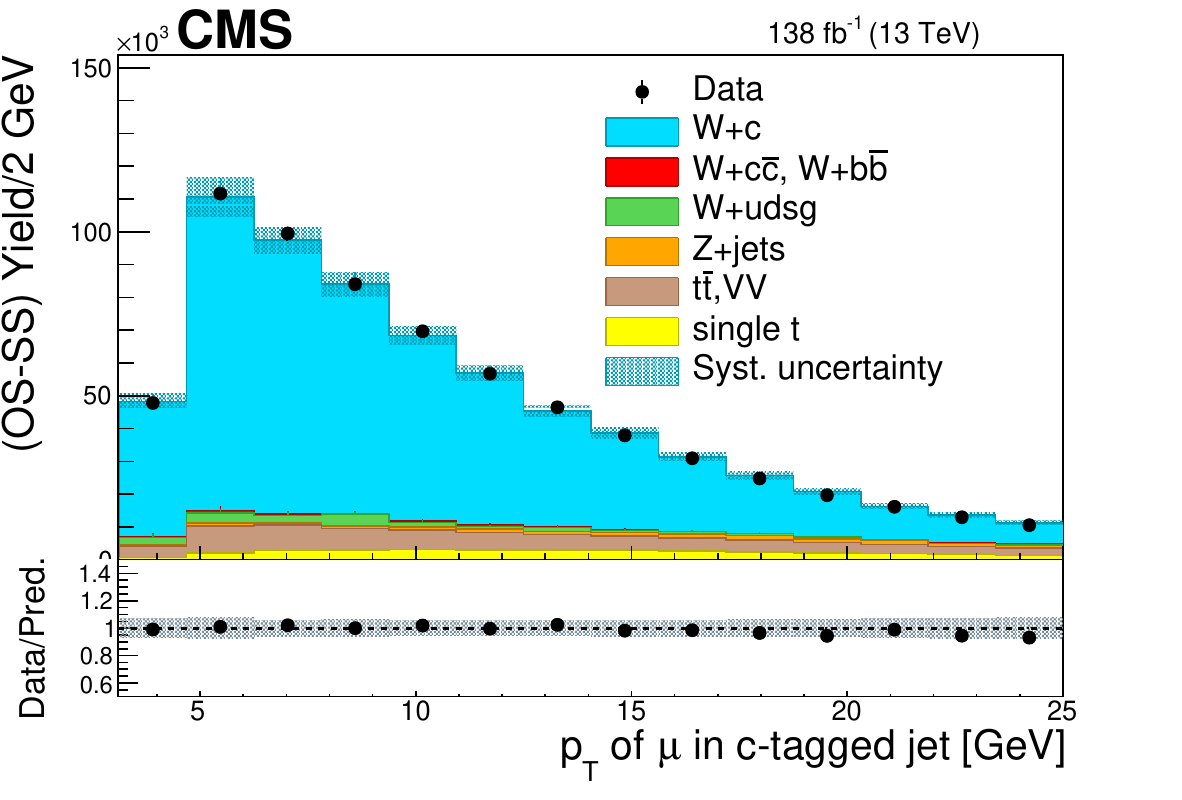}
\includegraphics[width=0.49\textwidth]{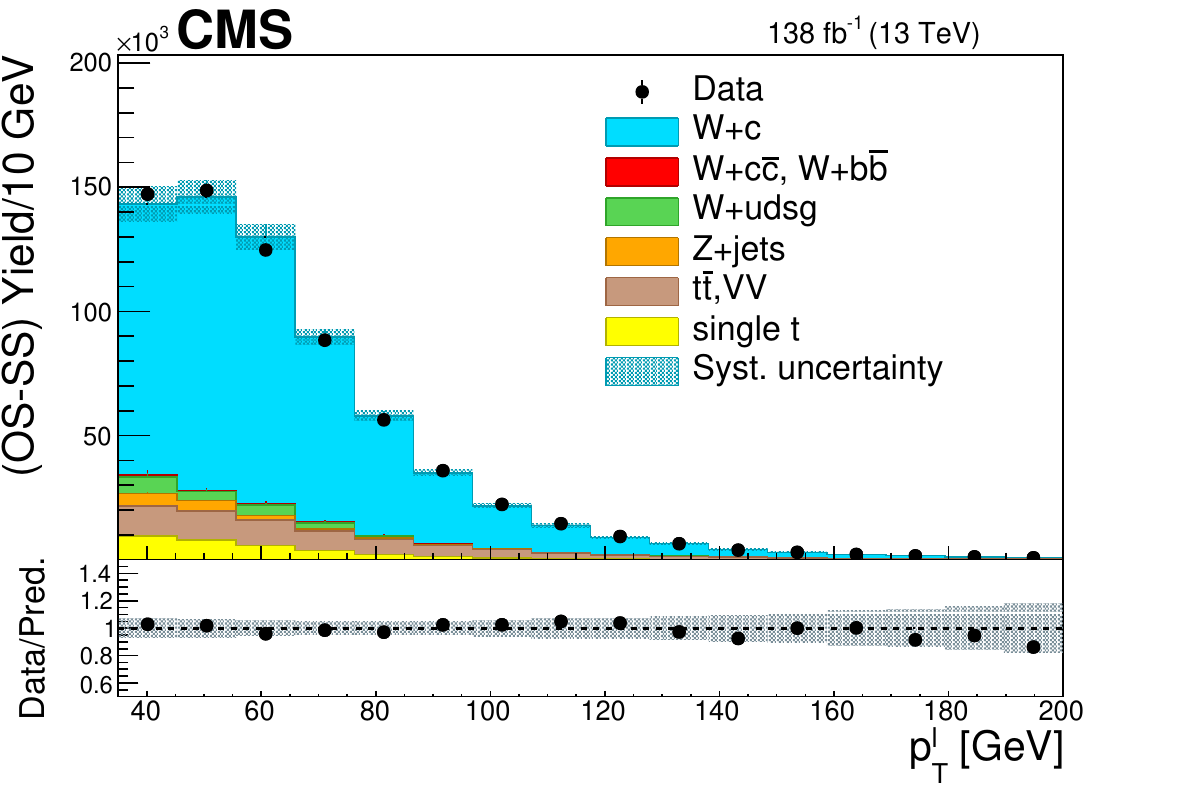}
\caption{Distributions  after OS-SS subtraction of the $\pt$
of the muon inside the $\cjet$ (\cmsLeft) and the  $\pt$ of the lepton from the \PW decay (\cmsRight) for events in the SL sample, summing up the contributions of the \PW boson decay channels to electrons and muons.
The contributions of the various processes are estimated with the simulated samples.
The statistical uncertainty in the data is smaller than the size of the data dots. The hatched areas represent the sum in quadrature of statistical and systematic uncertainties in the MC simulation. The ratio
of data to simulation is shown in the lower panels. The uncertainty band in the ratio includes the statistical uncertainty in the data, and the statistical and systematic uncertainties in the MC simulation.
}
\label{fig:muon_in_jet_pt}
\end{figure}

Figure~\ref{fig:muon_in_jet_pt} shows the $\pt$ distributions of the muon inside the jet (\cmsLeft) and of the lepton from the \PW decay (\cmsRight), for events in the selected SL sample, after the background normalization corrections described above. The simulations agree with the data within uncertainties.

\subsubsection{Event selection in the SV channel~\label{sec:Wsel_SV}}

An independent $\Wc$ sample is selected by looking for secondary decay vertices of charmed hadrons within the reconstructed jets. 
Displaced SVs are reconstructed with either the simple secondary vertex (SSV)~\cite{CMS-PAPER-BTV-12-001} or the
inclusive vertex finder (IVF)~\cite{Khachatryan:2011wq, Chatrchyan:2013zja} algorithms.
Both algorithms follow the adaptive vertex fitter technique~\cite{Adaptive_Vertex} to  construct an SV, but differ in the track selection used. 
The SSV algorithm takes as input the tracks constituting the jet, whereas the IVF algorithm starts from a displaced track with respect to the PV
(seed track) and tries to build a vertex from nearby tracks in terms of their separation distance in three dimensions and their angular separation
around the seed track. The IVF vertices are then associated to the closest jet in a cone of $\Delta R=0.3$. 
Both SSV and IVF vertices always start with input tracks with a minimum $\pt$ of $1\GeV$ to minimize the effects from poorly reconstructed tracks. 
Vertices reconstructed with the IVF algorithm are considered first. 
If no IVF vertex is found, SSV vertices are searched for, thus providing additional event candidates (about 3\%).
If more than one SV is reconstructed within a jet, the one with the highest $\pt$, computed from its associated tracks, is selected.
If there are several jets with an SV, only the SV associated to the jet of highest  $\pt$ is selected.

At least three tracks must be associated with an SV for it to be considered.
This requirement largely reduces the contamination of jets coming from the hadronization of light-flavor quarks ($\PQu$, $\PQd$, and $\PQs$) or gluons. It also reduces the systematic uncertainty associated with the SV reconstruction efficiency. To ensure that the SV is well separated from the PV, we require the displacement significance, defined as the three dimensional distance between the PV and SV, divided by its uncertainty, to be larger than 8. This requirement suppresses the $\Wlight$ background contribution below 1\%.  

To classify the event as OS or SS, we measure the sign of the charge of the charm quark produced in the hard interaction. For charged charm hadrons, the sum of the charges of the decay products reflects the charge of the \PQc quark. For neutral charm hadrons, the charge of the closest hadron produced in the fragmentation process can indicate the charge of the \PQc quark~\cite{chargedeterminationreference, chargedeterminationreference2}. Hence, we assign a charge equal to the sum of the charges of the particle tracks associated with the SV. If the SV charge is zero, we assume the charge of the track that is closest in angular separation to the SV. We only consider PV tracks with $\pt>0.3\GeV$ and within an angular separation from the SV direction of 0.1 in the $(\eta, \phi)$ space. If nonzero charge cannot be assigned, the event is rejected. 
According to the simulation, the charge assignment procedure provides a nonzero charge for 99\% of the selected SVs, and the sign of the charge is correctly assigned in 83\% of the cases. 

The modeling of the SV charge assignment in the simulation has been validated with data. Events passing both the SL and SV selection criteria are used to compare the charges of the muon inside the jet and the SV. In 95\% of these events the charges agree. The difference in the charge assignment efficiency between data and simulation, around 1\%, is taken as a systematic uncertainty in the cross section measurements, as detailed in Section~\ref{sec:syst_uncert}.

The SV reconstruction efficiency in the simulation is calibrated using data. The events of the SL sample are used to compute
data-to-simulation scale factors for the efficiency of charm identification through the reconstruction of an SV~\cite{EPJC78, CMS-PAPER-BTV-16-002}.
The fraction of events in the SL sample with an SV is computed for data and simulation, and the ratio
of data to simulation is applied as a scale factor to the simulated $\Wc$ signal events in the SV sample. 
The calculated scale factor is $0.93 \pm 0.03$, where the uncertainty accounts for statistical and systematic effects.
The systematic uncertainty includes contributions from uncertainties in the pileup description, JES and JER,
lepton efficiencies, background subtraction, and modeling of charm production and decay fractions in the simulation.

Table~\ref{yieldsSV} shows the event yields in the $\Wen$ and $\Wmn$ channels after the selection requirements and $\OSSS$ subtraction.\
The background yields, as estimated with the simulations, are also included. The contributions of the backgrounds were rescaled using the normalization factors described in Section~\ref{sec:Wsel_SL}.
The signal and background composition of the selected sample, as predicted by the simulation, are shown in Table~\ref{fractiontable}. The purity of signal $\Wc$ events is above 80\%. 
The dominant backgrounds come from $\ttbar$ (8\%) and single top (9\%) production.

\begin{table}[htbp]
 \centering
  \topcaption{Data and background event yields (with statistical uncertainties) after selection and \OSSS subtraction for the SV channels (electron and muon \PW decay modes).}
  \label{yieldsSV}  \renewcommand{\arraystretch}{1.2}
  \begin{tabular}{ccc}
    SV  channel  & Data  & Background  \\
    \hline
    $\Wen$ & 338\,504 $\pm$ 1717 & 60\,565 $\pm$ 1577  \\
    $\Wmn$ & 494\,264 $\pm$ 1876 & 94\,356 $\pm$ 2002 \\
  \end{tabular}
\end{table}

\begin{table*}[htbp]
 \centering
  \topcaption{ Simulated signal and background composition (in percentage) of the SV sample after selection and \OSSS subtraction. The $\WQQ$ stands for the sum of the contributions of \Wcc~and \Wbb.}
  \label{fractiontable} 
  \renewcommand{\arraystretch}{1.2}
  \begin{tabular}{lccccccc}
   SV channel & \Wc  & \WQQ  & \Wlight  & $\Zj$  & \ttbar & Single \PQt & $\PV\PV$ \\
   \hline  
   $\Wen$ & 82.1 $\pm$ 0.8 & 0.7 $\pm$ 0.4 & 1.0 $\pm$ 0.6 & 0.1 $\pm$ 0.2 & 7.2 $\pm$ 0.1 & 8.4 $\pm$ 0.1 & 0.5 $\pm$ 0.1 \\
   $\Wmn$ & 80.9 $\pm$ 0.6 & 0.7 $\pm$ 0.3 & 0.5 $\pm$ 0.4 & 0.5 $\pm$ 0.2 &  8.0 $\pm$ 0.1 & 8.9 $\pm$ 0.1 & 0.5 $\pm$ 0.1 \\
  \end{tabular}
\end{table*}

The event selection requirements are summarized in Table~\ref{cutssummary} for the four selection channels of the analysis, the \PW boson decay channels to both electrons or muons, and the SL and SV charm identification channels.

\begin{table*}[htbp]
 \centering
  \topcaption{Summary of the selection requirements for the four selection channels of the analysis.}
  \label{cutssummary}
  \begin{tabular}{l|c|c|c|c}
   & SL &  SL &  SV & SV \\ 
   & $\Wen$ & $\Wmn$ & $\Wen$ & $\Wmn$ \\
   \hline
   Lepton $\pt^{\ell}$ & \multicolumn{4}{c}{${>}35\GeV$} \\
   Lepton $\abs{\eta^{\ell}}$ & \multicolumn{4}{c}{${<}2.4$} \\
   Lepton isolation $\IComb/\pt^{\ell}$ & \multicolumn{4}{c}{${<}0.15$} \\
   Transverse mass $\MT$ & \multicolumn{4}{c}{${>}55\GeV$} \\
   Jet $\pt^{\jet}$ & \multicolumn{4}{c}{${>}30\GeV$} \\
   Jet $\abs{\eta^{\jet}}$ & \multicolumn{4}{c}{${<}2.4$} \\
   $\Delta R ({\text{jet}},\ell)$ & \multicolumn{4}{c}{${>}0.4$} \\
   \hline
   Muon in jet $\pt^{\PGm}$ & \multicolumn{2}{c|}{${<}25\GeV$} & \multicolumn{2}{c}{} \\
   Muon in jet $\pt^{\PGm}/\pt^{\jet}$ & \multicolumn{2}{c|}{${<}0.6$} & \multicolumn{2}{c}{} \\
   Muon in jet $\abs{\eta^{\PGm}}$ & \multicolumn{2}{c|}{${<}2.4$} & \multicolumn{2}{c}{} \\
   \hline
   Muon in jet IPS & & ${>}2$ & \multicolumn{2}{c}{} \\
   \multirow{2}{*}{Muon in jet $m_{\PGm\PGm}$} & & ${>}12\GeV$ \& & \multicolumn{2}{c}{} \\
    & & ${\notin}[70,110\GeV{}]$  & \multicolumn{2}{c}{} \\
   \hline
   SV number of tracks & \multicolumn{2}{c|}{} & \multicolumn{2}{c}{${>}2$}\\
   SV displacement significance & \multicolumn{2}{c|}{} & \multicolumn{2}{c}{${>}8$}\\
   SV charge & \multicolumn{2}{c|}{} & \multicolumn{2}{c}{${\neq}0$}\\
  \end{tabular}
\end{table*}

\begin{figure}[!tb]
\centering
\includegraphics[width=0.49\textwidth]{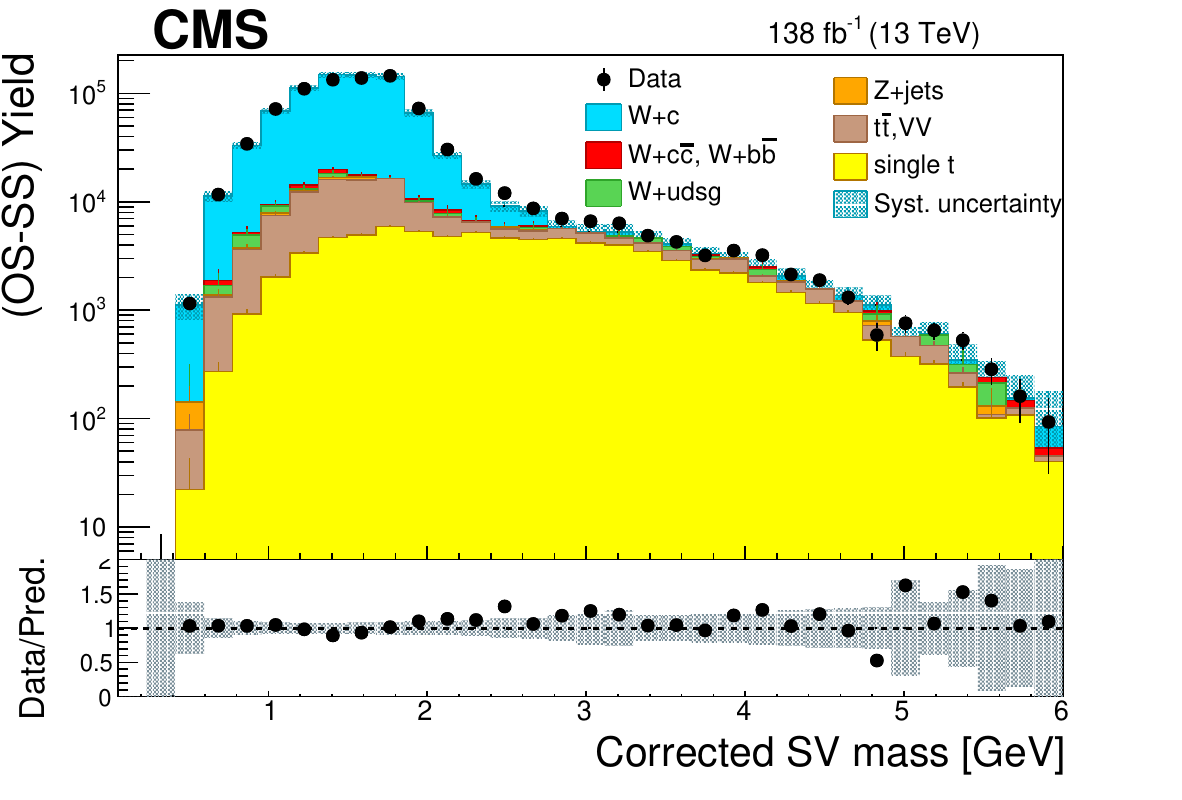}
\includegraphics[width=0.49\textwidth]{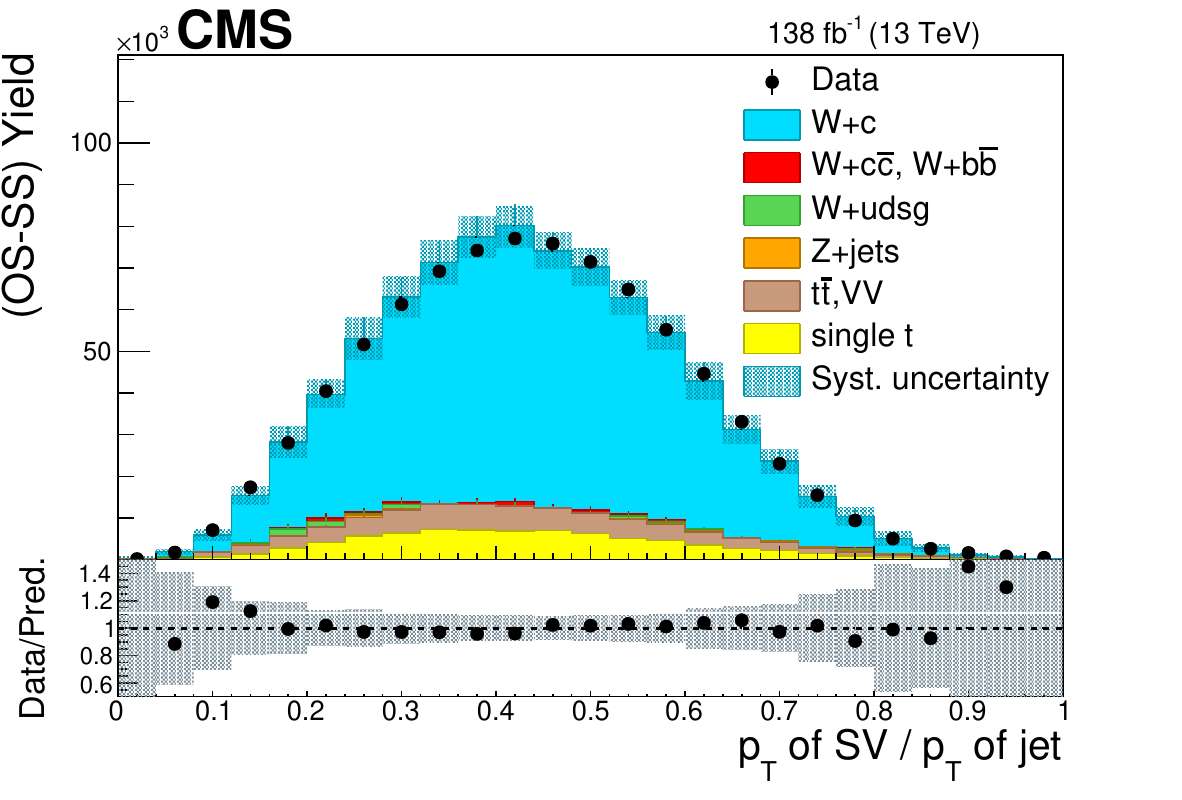}
\caption{Distributions after \OSSS subtraction of the corrected SV mass (\cmsLeft) and SV transverse momentum divided by the jet transverse momentum (\cmsRight) for events in the SV sample, summing
up the contributions of the \PW boson decay channels to electrons and muons. The contributions from all processes
are estimated with the simulated samples. The statistical uncertainty in the data is smaller than the size of the data dots for most of the data points. The hatched areas represent the sum in quadrature of statistical and systematic uncertainties in the
 MC simulation. The ratio of data to simulation is shown in the lower panels. The uncertainty band in the ratio includes the statistical uncertainty in the data, and the statistical and systematic uncertainties in the MC simulation.}
\label{fig:sv_in_jet_pt}
\end{figure}

Figure~\ref{fig:sv_in_jet_pt} shows the distributions, after \OSSS subtraction, of
the corrected SV mass and the SV transverse momentum divided by the jet transverse momentum, $\pt^{\text{SV}}$/$\pt^{\jet}$. The latter is a representative observable of the energy fraction of the charm quark carried by the charm hadron in the fragmentation process.
We define the corrected SV mass, $m_\text{SV}^\text{corr}$, as the invariant mass of all charged particles associated with the SV, assumed to be pions, $m_\text{SV}$, corrected for additional particles, either charged or neutral, that may have been produced but were not reconstructed~\cite{Aaij:2015yqa}: 
\begin{linenomath*}
\begin{equation*}
 m_\text{SV}^\text{corr} = \sqrt{m^2_\text{SV} + p^2_\text{SV} \sin^2 \theta}  + p_\text{SV} \sin \theta,
\end{equation*}
\end{linenomath*}
where $p_\text{SV}$ is the modulus of the vectorial sum of the momenta of all charged particles
associated with the SV, and $\theta$ is the angle between the momentum vector sum and the vector from the PV to the SV. The corrected SV mass is thus the minimum mass the long-lived hadron can have that is consistent with the direction of its momentum.

 The normalization of the single top quark background is fixed with data. Single top quark events populate the tail of the $ m_\text{SV}^\text{corr}$ distribution. A normalization factor of $1.5\pm0.2$ for the single top quark contribution was required to match data and simulation predictions. 
The same rescaling is applied to the SL and SV samples. 

\section{Systematic uncertainties~\label{sec:syst_uncert}}
The impact of various sources of uncertainty in the measurements presented in Section~\ref{sec:xsec_meas} is estimated by recalculating the cross sections 
with the relevant parameters varied up and down by one standard deviation of their uncertainties. 

The combined uncertainty in the trigger, reconstruction, and identification efficiencies for isolated leptons results in an uncertainty in the cross section measurements of about 2\% (1\%) for the $\Wen$ ($\Wmn$) channel. 
The uncertainty in the identification efficiency of nonisolated muons inside jets is approximately 3\%, according to dedicated studies with $\Zj$ events. This uncertainty affects only the SL channel.

The effects of the uncertainty in the JES and JER are assessed by 
varying up and down the $\pt$ values of jets with the corresponding uncertainty factors. The JES
and JER uncertainties are also propagated to $\ptvecmiss$. 
The resulting uncertainty in the cross section is about 2\% (1\%) for the SL (SV) channel. 
The uncertainty from a $\ptvecmiss$  mismeasurement in the event is estimated by varying within its uncertainty the contribution of the energy unassociated with reconstructed particle-flow objects. The effect in the cross section measurement is ${<}0.5\%$.
Uncertainties in the pileup modeling are calculated using a modified pileup profile obtained by changing the mean number of interactions by ${\approx}5\%$.
This variation covers the uncertainty in the $\pp$ inelastic cross section~\cite{CMS:2018mlc} and in the modeling of the pileup simulation. 
It results in less than 0.5\% uncertainty in the cross section measurements.

The integrated luminosities of the 2016, 2017, and 2018 data-taking periods are individually known with uncertainties in the 1.2--2.5\% range~\cite{CMS-LUM-17-003,CMS-PAS-LUM-17-004,CMS-PAS-LUM-18-002}, whereas the total 2016--2018 integrated luminosity has an uncertainty of 1.6\%. The improvement in precision arises from the (uncorrelated) time evolution of some systematic effects.

The uncertainty in the scale factor correcting the SV reconstruction efficiency in the simulation propagates into a systematic uncertainty of 3\% in the cross section. The uncertainty in the SV charge determination is estimated as the difference (1\%) in the rate obtained in data and simulation of correct SV charge assignment in the validation test described in Section~\ref{sec:Wsel_SV}.

Because of the dependence of the SV reconstruction efficiency on the SV displacement, we have evaluated the effect produced by an imperfect modeling of this observable by reweighting the SV displacement significance distribution of the simulation to match that of the data. The resulting uncertainty in the cross section measurement is 1--2\%. In addition, the stability of the results with the minimum SV displacement significance requirement was checked by changing the threshold from 8 to 7. The effect in the results is also at the 1\% level.  
 
The background contributions are evaluated with the simulations validated in data control samples, as discussed in Section~\ref{sec:Wsel_SL}. The uncertainty in the predicted background levels has an effect of 1\% in the cross section measurements.

The signal samples used for the acceptance and efficiency calculations were generated with \MADGRAPH+ \PYTHIAeight using the NNPDF3.0 and NNPDF3.1 PDF sets. The envelope of the systematic variations (replicas) of the nominal PDF is assumed to be the systematic uncertainty due to an imperfect knowledge of the PDFs, as recommended in Ref.~\cite{PDF4LHC}. The effect is approximately $1\%$. The statistical uncertainty in the determination of the selection efficiency using the simulated samples is 1\%, and is propagated as an additional systematic uncertainty.

In the signal and background modeling, no uncertainty is included in the simulation of higher-order terms in perturbative QCD (parton shower). The \OSSS subtraction technique removes the contribution to $\Wc$ production coming from charm quark-antiquark pair production, rendering the measurement insensitive to this effect. The uncertainty in the modeling of the hard process in the signal simulation is assessed by independently changing the QCD factorization and renormalization scales by factors of 0.5 and 2 relative to the nominal value. The resulting uncertainty in the cross section measurement is negligible. 

To estimate the effect produced by the uncertainties in the corrected values used in the simulation for the charm fragmentation and decay branching fractions~\cite{Lisovyi:2015uqa,PDG22}, we have varied those values within their uncertainties. The impact in the cross section measurements is 1--2\%, both for the fragmentation and decay branching fractions. 

The main systematic uncertainties are summarized in Table~\ref{syst_uncert} for the four selection channels of the analysis. Overall, the total systematic uncertainty in the $\Wc$ fiducial cross section is approximately $5\%$ in all channels.

\begin{table*}[htbp]
 \centering
  \topcaption{Summary of the main systematic uncertainties, in percentage of the measured fiducial cross section,  for the four selection channels of the analysis. }
  \label{syst_uncert} 
  \renewcommand{\arraystretch}{1.2}
  \begin{tabular}{lcccc}
    & SL &  SL &  SV & SV \\ 
    & $\Wen$ & $\Wmn$ & $\Wen$ & $\Wmn$ \\
   \hline
   Source & \multicolumn{4}{c}{Uncertainty [\%]} \\
   \hline
   Isolated lepton identification & 1.6 & 0.9 & 1.6 & 0.9 \\
   Jet energy scale and resolution & 2.0 & 2.0 & 1.0 & 1.0 \\
   Muon in jet identification & 3.0 & 3.0 & \NA & \NA \\
   SV reconstruction & \NA & \NA & 3.7 & 3.7 \\
   Charm fragmentation and decay & 2.6 & 2.6 & 2.4 & 2.4 \\
   PDF in MC samples & 1.0 & 1.0 & 1.0 & 1.0 \\
   Stat. uncert. selection efficiency & 0.9 & 1.2 & 0.9 & 0.8 \\
   Background contributions & 0.6 & 0.9 & 1.3 & 1.3 \\
   Integrated luminosity & 1.6 & 1.6 & 1.6 & 1.6  \\[\cmsTabSkip]

   Total & 5.2 & 5.1 & 5.4 & 5.2 \\
   \end{tabular}
\end{table*}

\section{Cross section measurements~\label{sec:xsec_meas}}
The $\Wc$ production cross section measurements are restricted to a phase space region that is close to the experimental fiducial volume with optimized sensitivity for the signal process.
Cross sections are measured inclusively within the fiducial phase space region and differentially as a function of
$\pt^{\ell}$ and $\abs{\eta^{\ell}}$.
Cross section measurements are performed independently in four different channels, the two charm identification SL and SV channels, and the two \PW boson decay channels. The four measurements are combined to improve the precision.

Measurements are unfolded to the particle and parton levels.
At both levels, the fiducial region is defined by a lepton at the generator level coming from the decay of a \PW boson with $\pt^{\ell}>35\GeV$ and $\abs{\eta^{\ell}} < 2.4$, together with a generator-level \PQc jet with $\pt^{\cjet} > 30\GeV$ and $\abs{\eta^{\cjet}} < 2.4$. 
The \OSSS subtraction is also applied at generator level. This removes the charge-symmetric contributions that mostly originate from gluon splitting into a charm quark-antiquark pair.  
The \PQc jet must be well separated from the lepton by an angular distance $\Delta R (\cjet,\ell)>0.4$. Jets at the generator level are clustered using the anti-$\kt$ jet algorithm with a distance parameter $R = 0.4$. At the particle level, jets are formed using generator particles produced after the hadronization process. At the parton level, jets are constructed from the hard interaction partons.    

For all channels under study, the $\Wc$ cross section is determined using the following expression:
\begin{linenomath*}
\begin{equation}
\SWc = \frac{Y_{\text{sel}}-Y_{\text{bkg}}}{\mathcal{C} \, \mathcal{L}},
\label{eq:W_c_data}
\end{equation}
\end{linenomath*}
where $Y_{\text{sel}}$ is the selected \OSSS event yield, and $Y_{\text{bkg}}$ the background yield in data after \OSSS subtraction, estimated from simulation and normalized using the data control samples described in Section~\ref{sec:Wjetssel}. $\mathcal{L}$ is the integrated luminosity of the data sample.

The factor $\mathcal{C}$ corrects for acceptance and efficiency losses in the selection process of  $\Wc$ events produced in the fiducial region at the generator level. 
It also subtracts the contributions from $\Wc$ events outside the kinematic region of the measurements and from $\Wc$ events with $\Wtn$, $\PGt\to\Pe + X$ or $\PGt\to\PGm + X$. 
It is calculated, using the sample of simulated signal events, as the ratio between the event yield of the selected $\Wc$ sample (according to the procedure 
described in Sections~\ref{sec:Wsel_SL} and~\ref{sec:Wsel_SV} and after  \OSSS subtraction) 
and the number of \OSSS $\Wc$ events satisfying the phase space definition at the generator level. Independent correction factors $\mathcal{C}$ are computed at the particle and parton levels, and for the four selection channels.

\subsection{Measurements at the particle level~\label{sec:xsec_particle}}
Cross section measurements, unfolded to the particle level, are presented in this section.  The fiducial $\Wc$ production cross section measurements 
computed with Eq.~(\ref{eq:W_c_data}) for the four channels separately are shown in Table~\ref{table:Cross_sectionsall}, together with the event yields and the $\mathcal{C}$ correction factors. The different $\mathcal{C}$ values reflect the different reconstruction and selection efficiencies in the four channels. In the SL channel, less than 5\% of the signal charm hadrons generated in the fiducial region of the analysis produce a muon in their decay with enough momentum to reach the muon detector and get reconstructed. Similarly, in the SV channel, less than 5\% of the events with a charm hadron decay remain after SV reconstruction, SV charge assignment, and \OSSS subtraction. The remaining inefficiency, accounted for in the $\mathcal{C}$ correction factor, is due to the selection requirements of the samples.

\begin{table}[htbp]
 \centering
 \caption{Measured production cross sections $\SWc$ unfolded to the particle level 
in the four selection channels 
together with statistical (first) and systematic (second) uncertainties. The acceptance times efficiency values ($\mathcal{C}$) are also given. 
}
\label{table:Cross_sectionsall}
 \begin{tabular}{ccc}
Channel &  ${\cal C} (\%)$ & $\SWc$ [pb] \\
 \hline
 $\Wen$, SL &  1.568 $\pm$ 0.014 $\pm$ 0.077  & 158.7 $\pm$ 0.6  $\pm$ 8.3  \\ 
 $\Wmn$, SL  &  0.946 $\pm$ 0.011 $\pm$ 0.044 & 149.8 $\pm$ 0.7  $\pm$ 7.7  \\ 
 $\Wen$, SV  &  1.389 $\pm$ 0.013 $\pm$ 0.068 & 145.0 $\pm$ 0.9  $\pm$ 7.6  \\ 
 $\Wmn$, SV  & 1.966 $\pm$ 0.015 $\pm$ 0.093  & 147.4 $\pm$ 0.7  $\pm$ 7.5  \\ 
 \end{tabular} 
 \end{table} 

Results obtained for the $\Wc$ cross sections in the four different channels are consistent within the uncertainties, and  are combined
using the best linear unbiased estimator method~\cite{BLUELyons} that takes into account individual uncertainties and their correlations. 
Systematic uncertainties arising from a common source and affecting several measurements are considered as fully correlated.
In particular, all systematic uncertainties are assumed fully correlated between the electron and muon channels, except those related to the lepton reconstruction. The $\chi^2$ of the combination is 4.8 (three degrees of freedom), corresponding to a p-value of 0.19. The combined measured cross section unfolded to the particle level is:
\begin{linenomath*}
  \begin{equation*}
  \begin{aligned}
      \sigma(\noppWc) & = 148.7 \pm 0.4 \stat \pm 5.6 \syst \unit{pb}. 
  \end{aligned}
  \end{equation*}
\end{linenomath*}

\begin{figure}[htbp!]
\centering
\includegraphics[width=0.5\textwidth]{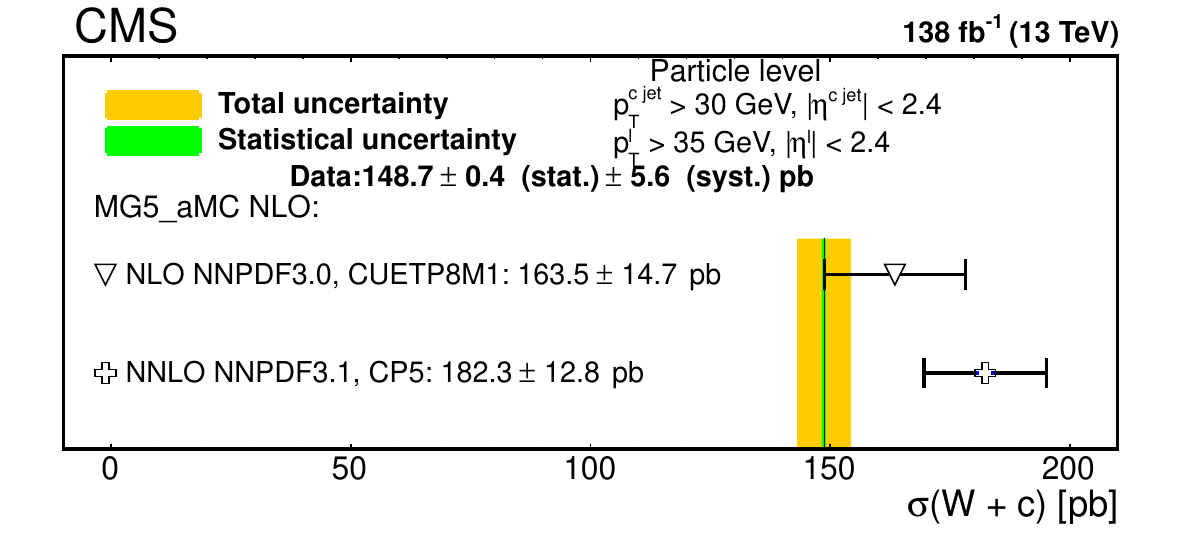}
\caption{Comparison of the measured fiducial $\SWc$ cross section unfolded to the particle level with the predictions from the \MGvATNLO simulation using two different PDF sets (NLO NNPDF3.0 and NNLO NNPDF3.1). Two different tunes (CUETP8M1 and CP5) for the parton showering, hadronization and underlying event modeling  in \PYTHIAeight are also used. Horizontal error bars indicate the total uncertainty in the predictions.}
\label{fig:summary_particle}
\end{figure}

Measurements are compared with the predictions of the \MGvATNLO MC generator, as shown in  Fig.~\ref{fig:summary_particle}. 
In the predictions, two different NNPDF PDF sets (versions 3.0 and 3.1)  are used. 
The two predictions differ as well in the tune used in \PYTHIAeight for the parton showering, hadronization, and underlying event modeling (CUETP8M1 and CP5).
The predicted cross sections are about 10\% (using NLO NNPDF3.0) and 20\% (NNLO NNPDF3.1) higher than the measured value, with relative uncertainties close to 10\%.  
The uncertainty associated with the MC predictions includes the uncertainties associated with the renormalization and factorization scales, as well as the uncertainty related to the PDFs used in the simulation. The scale uncertainties are estimated using a set of weights provided by the generator that corresponds to independent variations of the scales by factor of 0.5, 1, and 2. The prediction is obtained for all combinations (excluding the cases where one scale is reduced and the other is increased at the same time) and their envelope is quoted as the uncertainty. The uncertainty in the PDFs is estimated using different Hessian eigenvectors of each PDF set.

The $\SWc$ production cross section is also measured differentially as a function of $\abs{\eta^\ell}$ and $\pt^\ell$. The total sample is divided into subsamples according to the value of $\abs{\eta^\ell}$ or $\pt^\ell$,
and the cross section is computed using Eq.~(\ref{eq:W_c_data}).
The binning of the differential distributions is chosen such that each bin is sufficiently populated to perform the measurement. 
Event migration between neighbouring bins caused by detector resolution effects is evaluated with the simulated signal sample and is negligible. Measurements in the four channels are combined assuming that systematic uncertainties are fully correlated among bins of the differential distributions.

Systematic uncertainties in the differential $\SWc$ cross section measurements are in the range of 4--6\%.
The main sources of systematic uncertainty, as discussed in Section~\ref{sec:syst_uncert}, are related to the charm hadron fragmentation and decay fractions in the simulation (2\%), and the efficiency of identifying an SV or a muon inside a jet (3\%).

\begin{figure}[htbp!]
\centering
     \includegraphics[width=0.49\textwidth]{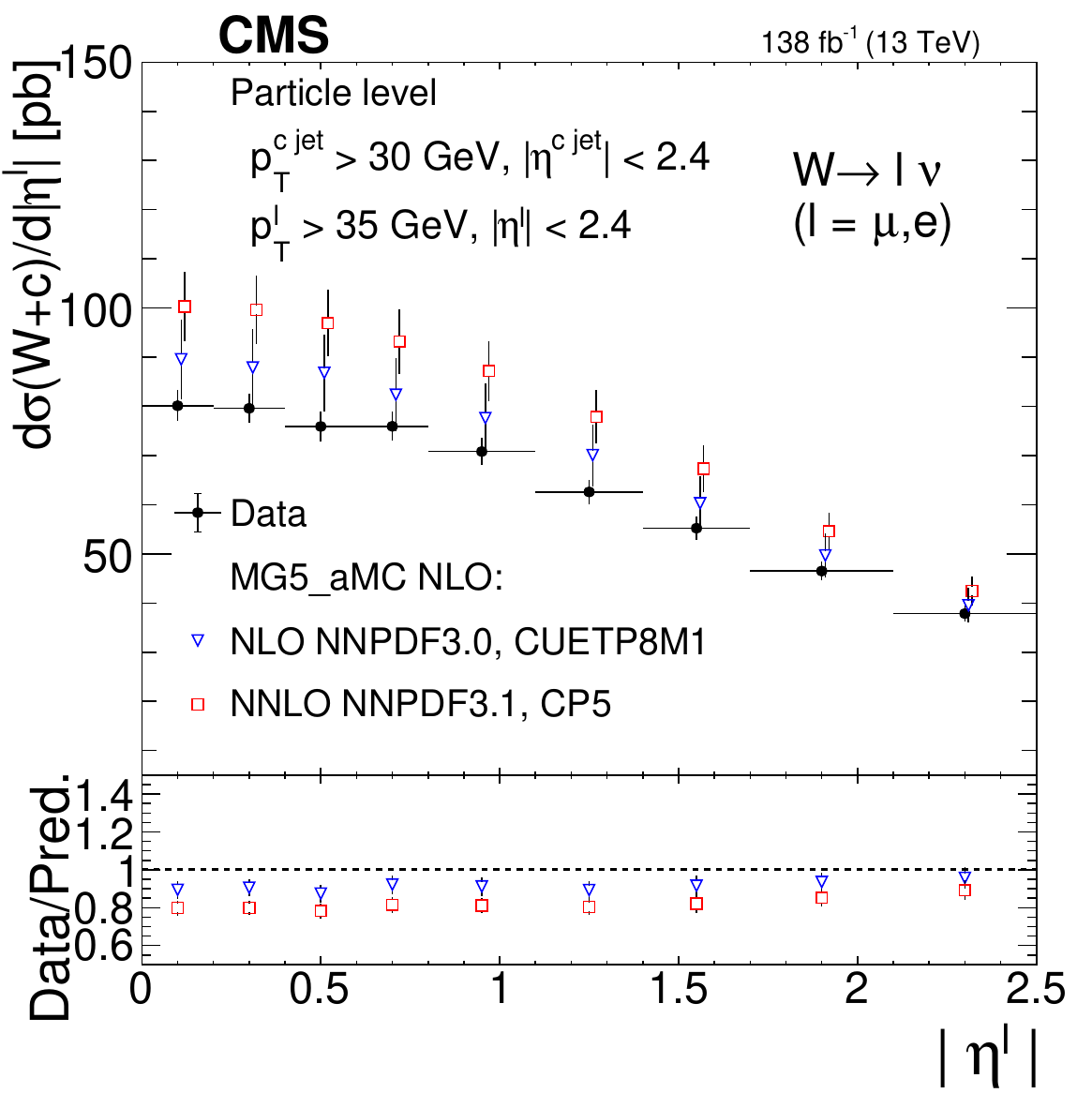}
     \includegraphics[width=0.49\textwidth]{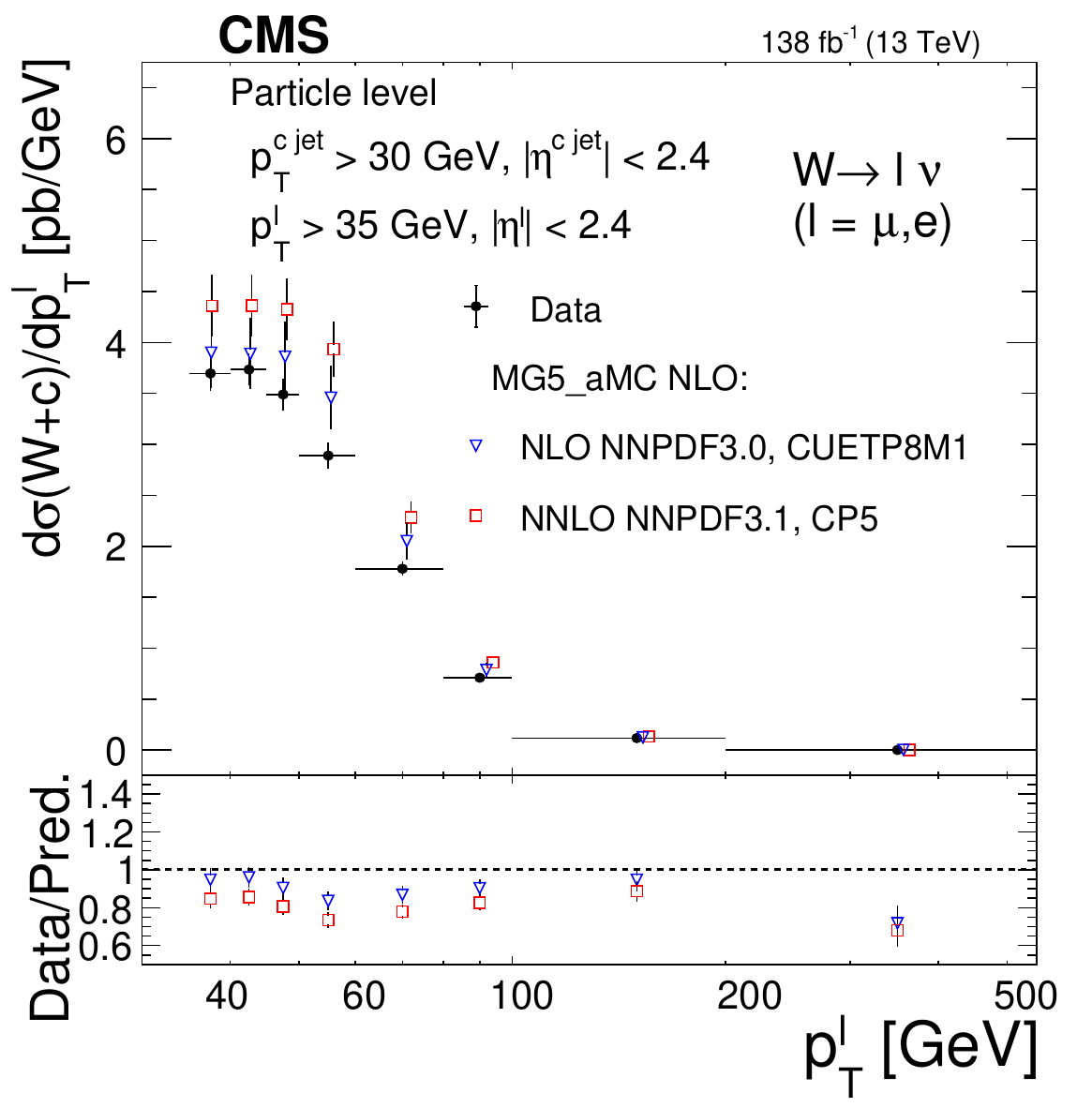}
    \caption{Measured differential cross sections $\SWcdifflineeta$ (\cmsLeft) 
             and $\SWcdifflinept$ (\cmsRight) unfolded to the particle level, compared with the predictions of the \MGvATNLO simulation. Two different PDF sets (NLO NNPDF3.0 and NNLO NNPDF3.1) are used. Error bars on data points include statistical and systematic uncertainties. Symbols showing the theoretical expectations are slightly displaced in the horizontal axis for better visibility. The ratios of data to predictions are shown in the lower panels. The uncertainty in the ratio includes the uncertainties in both data and prediction.}
\label{fig:Sc_w_th}
\end{figure}

The $\SWc$ differential cross section as a function of $\abs{\eta^{\ell}}$ and $\pt^\ell$, obtained after the combination of the measurements in the SL, SV, electron, and muon channels, 
is shown in Fig.~\ref{fig:Sc_w_th}, 
compared with the predictions from the \MGvATNLO simulation. 
Observed shape differences are within 10\%.

\subsection{Measurements at the parton level~\label{sec:xsec_parton}}

The measurements are also unfolded to the parton level, including an additional correction to account for the \PQc quark fragmentation and hadronization processes. Results of the fiducial cross sections in the four selection channels are presented in Table~\ref{table:Cross_sectionsall_parton}. The combination of the measurements is:
\begin{linenomath*}
  \begin{equation*}
      \sigma(\noppWc)  =  163.4 \pm 0.5\stat \pm 6.2\syst\unit{pb}. 
  \end{equation*}
\end{linenomath*}

The fiducial cross section measured at the parton level is expected to be slightly larger than that at the particle level.
During the hadronization and jet clustering processes, the momentum of the \PQc quark gets smeared and biased towards slightly smaller values. A fraction of charm quarks near the $\pt^{\PQc}>30\GeV$ threshold of the fiducial region of the measurement do not result in \PQc jets with $\pt^{\cjet} > 30\GeV$. On the other hand, a number of $\Wc$ events with a \PQc quark with $\pt^{\PQc} < 30\GeV$ get reconstructed with a generator level jet with  $\pt^{\cjet} > 30\GeV$. The net effect is the the observed reduction of about 10\% of the cross section at the particle level. 

\begin{table}[htbp]
 \centering
 \caption{Measured production cross sections $\SWc$ unfolded to the parton level in the four selection channels together with statistical (first) and systematic (second) uncertainties. The acceptance times efficiency values ($\mathcal{C}$) are also given.
}
\label{table:Cross_sectionsall_parton}
 \begin{tabular}{ccc}
Channel &  $\mathcal{C} (\%)$ & $\SWc$ [pb] \\
 \hline
 $\Wen$, SL  &  1.419 $\pm$ 0.012 $\pm$ 0.069 & 175.3 $\pm$ 0.7  $\pm$ 9.2  \\
 $\Wmn$, SL  &  0.856 $\pm$ 0.010 $\pm$ 0.040 & 165.4 $\pm$ 0.8  $\pm$ 8.5  \\
 $\Wen$, SV  &  1.261 $\pm$ 0.012 $\pm$ 0.062 & 159.6 $\pm$ 1.0  $\pm$ 8.4  \\
 $\Wmn$, SV  &  1.786 $\pm$ 0.014 $\pm$ 0.084 & 162.3 $\pm$ 0.8  $\pm$ 8.2  \\
 \end{tabular}
 \end{table}

{\tolerance=800 
The measurements unfolded to the parton level are compared with analytical calculations of $\Wc$ production. We have used the $\MCFM$ 9.1 program~\cite{MCFM9} to evaluate the cross section predictions in the phase space of the analysis: $\pt^{\ell}>35\GeV$, $\abs{\eta^{\ell}}<2.4$, $\pt^{\cjet}>30\GeV$, and $\abs{\eta^{\cjet}}<2.4$. Jets are clustered  in $\MCFM$ using the anti-$\kt$ jet algorithm with a distance parameter $R = 0.4$. The $\Wc$ process description is available in $\MCFM$ up to $\mathcal{O}({\alpS}^2)$ with a massive charm quark ($m_{\PQc} =1.5\GeV$). 
The contributions from gluon splitting into $\PQc\PAQc$ are not included. 
We have computed predictions for the following NLO PDF sets: MSHT20~\cite{MSHT2020}, CT18~\cite{CT18nlo}, CT18Z~\cite{CT18nlo}, ABMP16~\cite{ABMP16nlo}, NNPDF3.0~\cite{Ball:2014uwa}, and NNPDF3.1~\cite{NNPDF31nlo}. The LHAPDF6 library~\cite{LHAPDF} was used to access the PDF sets. All of the PDF sets were derived using strangeness-sensitive experimental data, including LHC \PW/\PZ and jet production cross section measurements. The NNPDF and MSHT20 sets additionally incorporate the CMS $\Wc$ production at $\sqrt{s}=~7\TeV$ data. CTA18Z differs from CTA18 because the former includes the ATLAS \PW/\PZ $7\TeV$ precision measurements~\cite{ATLAS-WZ-7TeV} leading to an enhancement of the strange PDF. The PDF parameterizations of the MSHT20 and NNPDF groups allow for strangeness asymmetry.    
\par}

The factorization and the renormalization scales are set to the value of the \PW boson mass~\cite{PDG22}. The uncertainty from missing higher perturbative orders is estimated by computing cross section predictions varying independently the factorization and renormalization scales to twice and half their nominal values, with the constraint that the ratio of scales is never larger than 2. The envelope of the resulting cross sections with these scale variations defines the theoretical scale uncertainty. The value in the calculation of the strong coupling constant at the energy scale of the mass of the $\PZ$ boson, $\alpS(m_{\PZ})$, 
is set to the recommended values by each of the PDF groups. 
Uncertainties in the predicted cross sections associated with $\alpS(m_{\PZ})$ are evaluated as half the difference in the predicted cross sections
evaluated with a variation of $\Delta(\alpS)=\pm 0.002$.  

\begin{figure}[htbp!]
\centering
\includegraphics[width=0.5\textwidth]{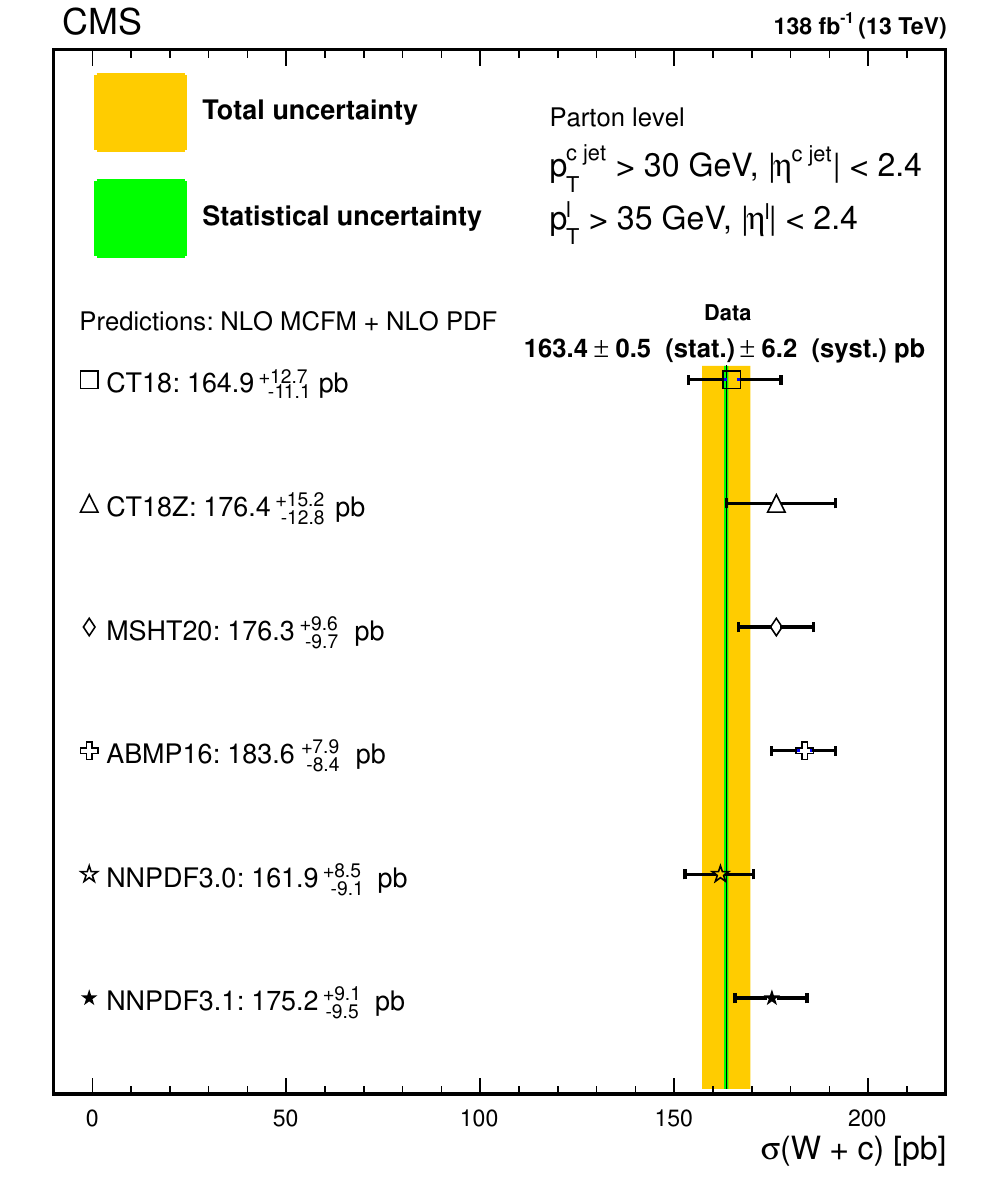}
\caption{
Comparison of the experimental measurement of $\SWc$, unfolded to the parton level, with the predictions from the NLO QCD $\MCFM$ calculations using different NLO PDF sets. Horizontal error bars indicate the total uncertainty in the predictions.
}
\label{fig:summary_parton}
\end{figure}

The theoretical predictions for the fiducial $\Wc$ cross section in the phase space of the measurements are summarized in Table~\ref{table:MCFM_Sc}. The central value of the prediction is provided together with the relative uncertainties arising from the PDF variations within each set, the choice of scales, and $\alpS$.
The size of the PDF uncertainties depends on the different input data and methodology
used by the various groups. In particular, they depend on the parameterization of the strange quark PDF and on the definition of the one standard deviation
uncertainty band. The maximum difference between the central values of the various PDF predictions is $\sim$10\%. 
This difference is comparable to the total uncertainty in each of the individual predictions.
Theoretical predictions are slightly larger than the measured cross section but are in agreement within the uncertainties, as depicted in Fig.~\ref{fig:summary_parton}.

\begin{table*}[htbp]
\centering
 \topcaption{Predictions for $\SWc$ production from $\MCFM$ at NLO in QCD for the phase space of the analysis.
For every PDF set, the central value of the prediction is given, together with the
 uncertainty as prescribed from the PDF set, and the  uncertainties associated with the scale variations and with the value of $\alpS$. The total uncertainty is given in the last column.
The last row in the table gives the experimental result presented in this paper.
}
\label{table:MCFM_Sc}
\renewcommand{\arraystretch}{1.2}
\begin{tabular}{cccccc}
PDF set   & $\SWc$ [pb] & $\Delta_{\text{PDF}}$ [pb] & $\Delta_{\text{scales}}$ [pb] & $\Delta_{\alpS}$ [pb] & Total uncert. [pb] \\ 
\hline
MSHT20  & $176.3$ & $^{+6.8}_{-6.3}$ & $^{+6.8}_{-7.4}$ & $\pm0.01$ & $^{+9.6}_{-9.7}$ \\[\cmsTabSkip]

CT18      & $164.9$ & $^{+11.1}_{-8.7}$ & $^{+6.1}_{-6.8}$ & $^{+0.9}_{-0.8}$ & $^{+12.7}_{-11.1}$  \\[\cmsTabSkip]

CT18Z  & $176.4$ & $^{+13.5}_{-10.5}$  & $^{+7.0}_{-7.4}$ & $^{+0.6}_{-0.5}$ & $^{+15.2}_{-12.8}$  \\[\cmsTabSkip]

ABMP16    & $183.6$ & $\pm 3.3$ & $^{+7.2}_{-7.8}$ & $^{+1.5}_{-0.9}$  & $^{+7.9}_{-8.4}$ \\[\cmsTabSkip]

NNPDF3.0   & $ 161.9$ & $\pm 6.2$   & $^{+5.8}_{-6.7}$ & $\pm0.01$  & $^{+8.5}_{-9.1}$ \\[\cmsTabSkip]

NNPDF3.1   & $ 175.2$ & $\pm 6.1$ & $^{+6.6}_{-7.3}$ & $\pm0.01$ & $^{+9.1}_{-9.5}$ \\[\cmsTabSkip]

\multicolumn{6}{c}{CMS: $163.4 \pm 0.5\stat \pm 6.2\syst\unit{pb}$}  \\
\end{tabular}
\end{table*}

\begin{figure}[htbp!]
\centering
     \includegraphics[width=0.49\textwidth]{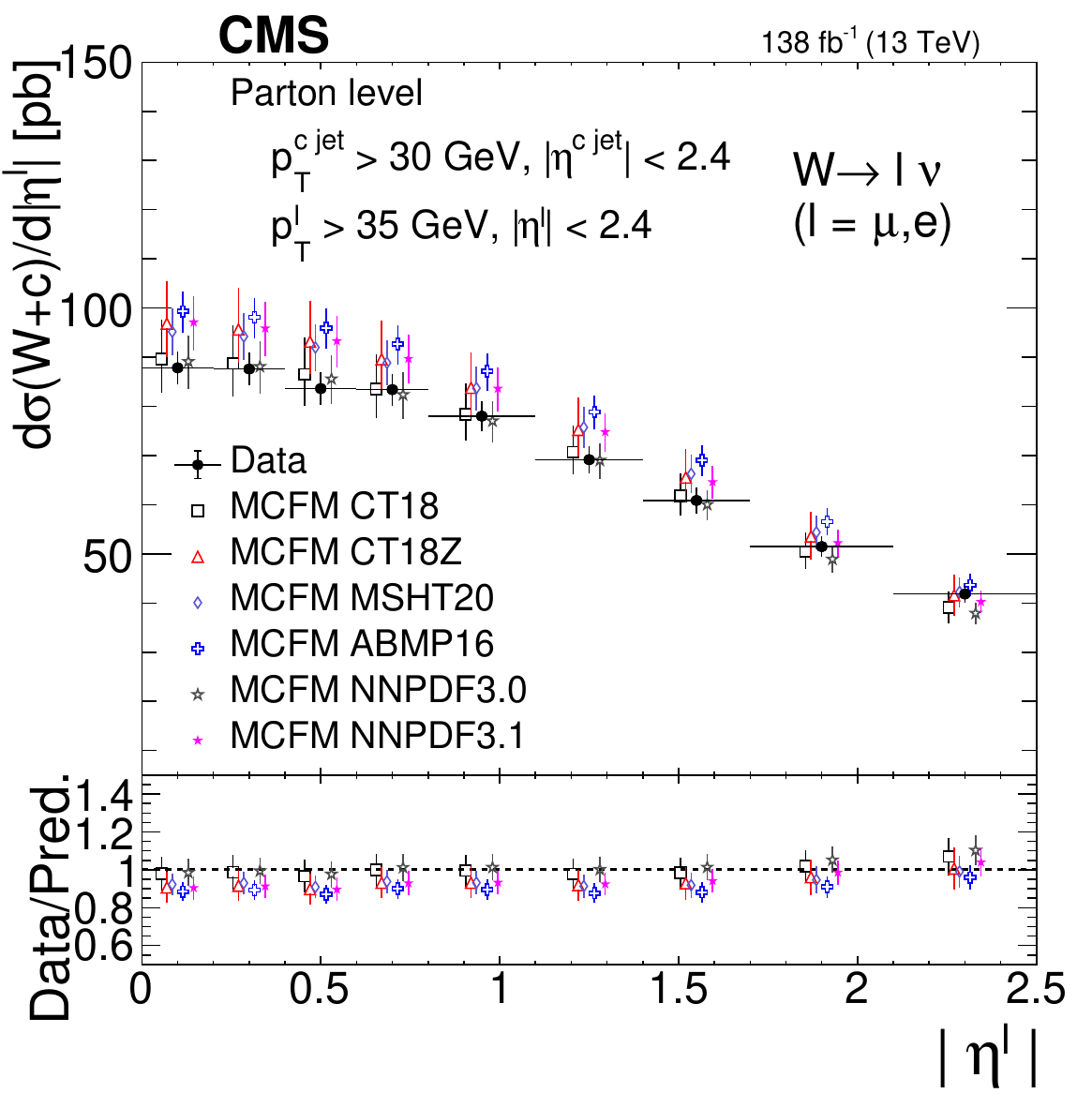}
     \includegraphics[width=0.49\textwidth]{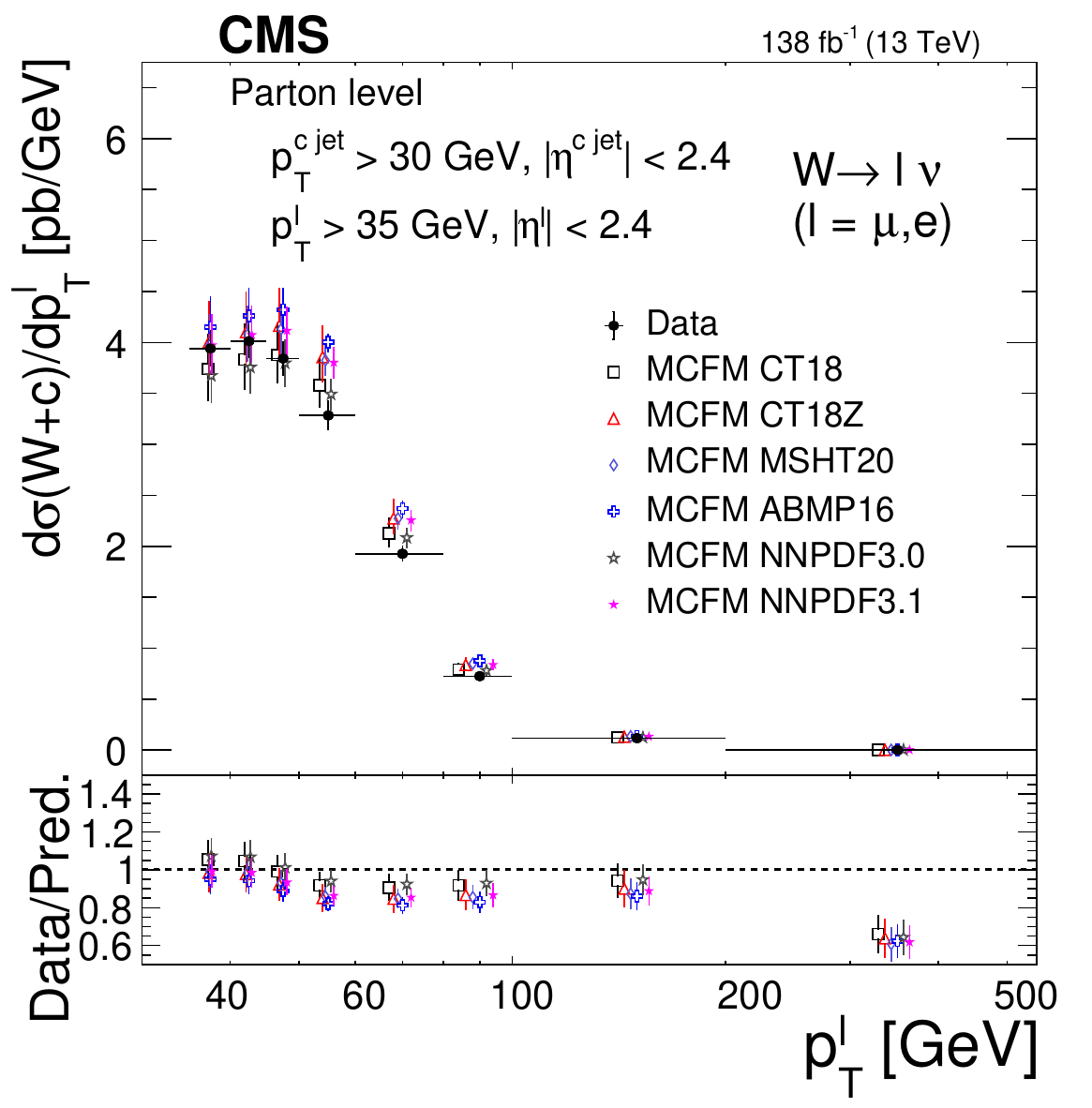}
    \caption{Measured differential cross sections $\SWcdifflineeta$ (\cmsLeft)
             and $\SWcdifflinept$ (\cmsRight) unfolded to the parton level,
compared with the predictions from the $\MCFM$ NLO calculations using different NLO PDF sets. Error bars on data points include statistical and systematic uncertainties. Symbols showing the theoretical expectations are slightly displaced in the horizontal axis for better visibility. The ratios of data to predictions are shown in the lower panels. The uncertainty in the ratio includes the uncertainties in both data and prediction.
}
\label{fig:Sc_w_th_parton}
\end{figure}

The predictions for the $\SWc$ production cross section, computed in intervals of $\abs{\eta^\ell}$ and $\pt^\ell$, are compared with the measured values in Fig.~\ref{fig:Sc_w_th_parton}. The predictions are generally consistent with the measurements within uncertainties, except for the highest $\pt^\ell$ bin.

\subsection{Measurements of the cross section ratio \texorpdfstring{$\SWpc/\SWmc$}{(W+ + cbar)/(W- + c)}~\label{sec:xsec-ratio}}
The cross section ratio  $\SWpc/\SWmc$ is measured in the four channels as the ratio of the \OSSS event yields in which the lepton from the \PW boson decay is positively or negatively charged:
\begin{linenomath*}
\begin{equation}
\Rcpm = \frac{\SWpc}{\SWmc} =\frac{Y_{\text{sel}}^{+}-Y_{\text{bkg}}^{+}}{Y_{\text{sel}}^{-}-Y_{\text{bkg}}^{-}}.
\label{eq:W_c_ratio}
\end{equation}
\end{linenomath*}

The \OSSS background contributions, $Y_{\text{bkg}}^{+}$ and $Y_{\text{bkg}}^{-}$, estimated with the simulations, are subtracted from the selected event yields $Y_{\text{sel}}^{+}$ and $Y_{\text{sel}}^{-}$. The statistical uncertainty in the background contributions in the four analysis channels is treated as a source of systematic uncertainty (0.5--0.8\%) in the cross section ratio.  

Most of the reconstruction and selection efficiencies cancel out in the measurement of the cross section ratio $\Rcpm$. Possible efficiency differences between positive and negative leptons and SVs are included as systematic uncertainties. We evaluate effects stemming from charge confusion and charge-dependent reconstruction efficiencies.

The probability of assigning the incorrect charge to a lepton is studied with data using $\Zll$ events reconstructed with SS or OS leptons. For the muons, the charge misidentification probability is negligible ($<10^{-3}$). For the electrons, the effect is around 1\% but propagates into a negligible uncertainty in the cross section ratio. The charge confusion rate for the SVs is significantly larger, 17\%, as described in Section~\ref{sec:Wsel_SV}. However, assuming that the charge confusion probability is the same for positive and negative SVs, the effect in the cross section ratio cancels out. 

Potential differences in the reconstruction efficiencies of positive and negative leptons or SVs are studied with the $\Wc$ MC simulation. Efficiency ratios are calculated independently for the four channels of the analysis and are consistent with unity within the statistical uncertainty (1.2--1.4\%). No corrections are made in the $\Rcpm$ measurements but the statistical uncertainties in the efficiency ratios are treated as systematic uncertainties.

The $\Rcpm$ measurements in the four channels are presented in Table~\ref{table:Cross_sectionspm}. The four measurements are combined considering as fully correlated the systematic uncertainties of electron, muon and SV reconstruction efficiencies affecting several channels. The $\chi^2$ of the combination is 3.3 (three degrees of freedom), corresponding to a p-value of 0.35.    
The combined cross section ratio measurement is:

\begin{linenomath*}
  \begin{equation*}
      \Rcpm   = 0.950 \pm 0.005\stat \pm 0.010\syst.
  \end{equation*}
\end{linenomath*}

The precision in the $\Rcpm$ measurement has been improved by a factor of two with respect to previous CMS measurements~\cite{CMS-PAPER-SMP-12-002, CMS-PAPER-SMP-18-013, CMS-PAPER-SMP-17-014}, leading to the most precise measurement of $\Rcpm$ to date. 

\begin{figure}[htbp!]
\centering
\includegraphics[width=0.5\textwidth]{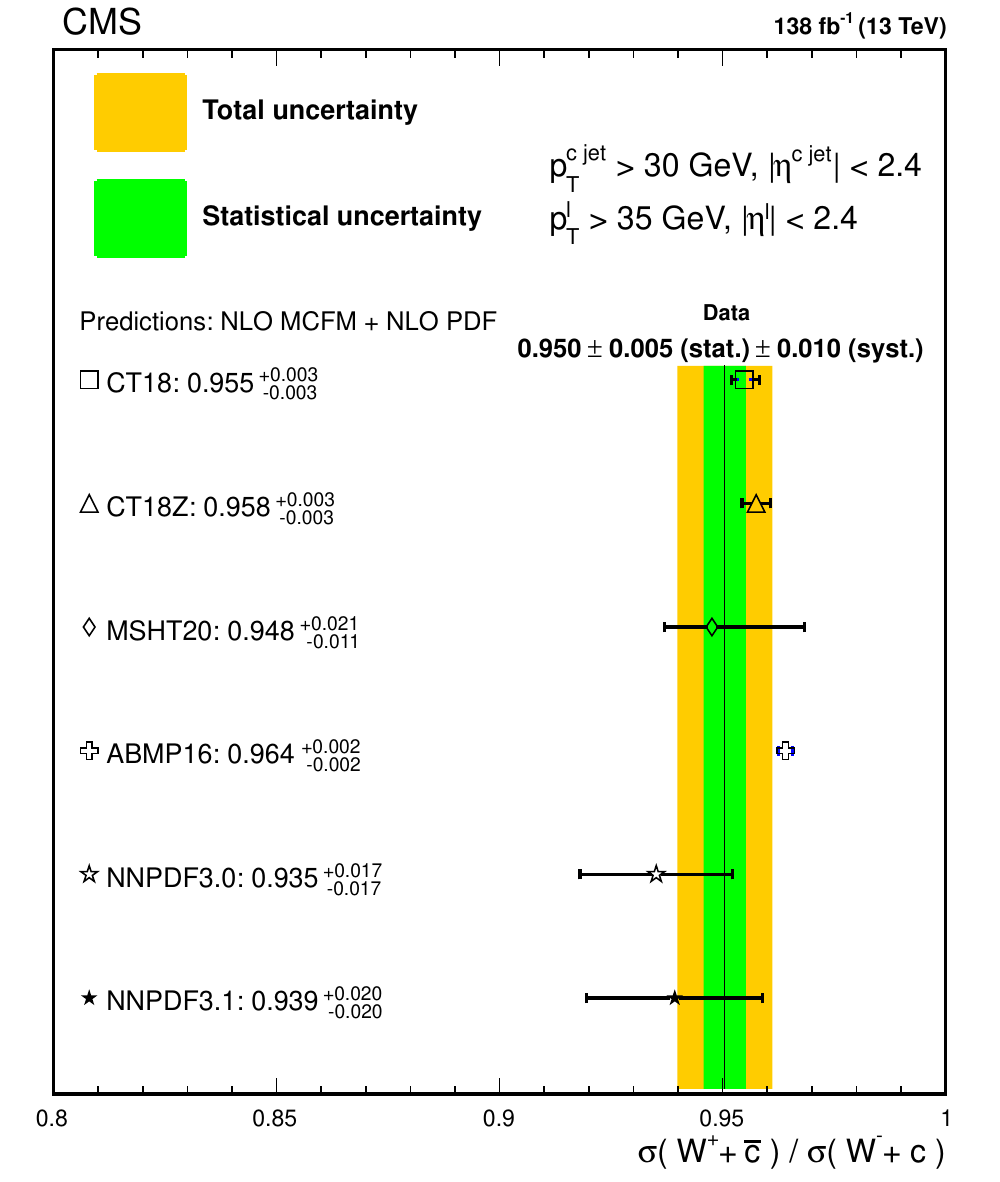}
\caption{
Comparison of the experimental measurement of $\Rcpm$ with
the NLO QCD $\MCFM$ calculations using different NLO PDF sets. Horizontal error bars indicate the total uncertainty in the predictions. 
}
\label{fig:Rcpm}
\end{figure}

In Fig.~\ref{fig:Rcpm} the $\Rcpm$ measurement is compared 
with the $\MCFM$ calculations using various PDF sets.  
Theoretical predictions for $\SWpc$ and $\SWmc$ are computed independently under the same conditions explained in Section~\ref{sec:xsec_parton} and for the same $\abs{\eta^{\ell}}$ and $\pt^{\ell}$ ranges used in the analysis. Expectations for $\Rcpm$ are derived from them and presented in Table~\ref{table:MCFM_Rcpm}.
All theoretical uncertainties are significantly reduced in the cross section ratio prediction. 

{\tolerance=9000 
The $\Rcpm$ observable is sensitive to a potential strangeness asymmetry in the proton but also to the down quark and antiquark asymmetry through the Cabibbo-suppressed down quark contribution to the $\Wc$ production. In the absence of strangeness asymmetry, as in the PDF sets CT18 and ABMP16, the predicted $\Rcpm$ value in the kinematical region of the analysis ranges from 0.955 to 0.964 with a small uncertainty of about 0.2\%. The predictions calculated using PDF sets that include strangeness asymmetry in the proton (MSHT20 and NNPDF) are about 2\% lower, ranging from 0.935 to 0.948 with a 2\% uncertainty as a result of the larger uncertainty associated with the difference between the strange quark and antiquark PDFs. 
Within experimental and theoretical uncertainties, the measured $\Rcpm$ value is consistent with both sets of predictions.   
\par}

\begin{table}[htbp] 
 \centering 
 \caption{Measured production cross section ratio $\Rcpm$ in the four selection channels. Statistical (first) and systematic (second) uncertainties are also given.
} 
\label{table:Cross_sectionspm} 
 \begin{tabular}{cc} 
Channel & $\Rcpm$\\ 
 \hline 
 $\Wen$, SL & 0.934 $\pm$ 0.006 $\pm$ 0.013 \\ 
 $\Wmn$, SL & 0.940 $\pm$ 0.006 $\pm$ 0.014 \\ 
 $\Wen$, SV & 0.961 $\pm$ 0.008 $\pm$ 0.013 \\
 $\Wmn$, SV & 0.974 $\pm$ 0.006 $\pm$ 0.015 \\ 
\end{tabular} 
\end{table}

\begin{table*}[htbp]
\centering
 \topcaption{Theoretical predictions for $\Rcpm$ calculated with $\MCFM$ at NLO.
 The kinematic selection follows the experimental requirements.
For every PDF set, the central value of the prediction is given, together with the
uncertainty as prescribed from the PDF set, and the uncertainties associated with the scale variations and with the value of $\alpS$. The total uncertainty is given in the last column.
The last row in the table gives the experimental result presented in this paper.
}
\label{table:MCFM_Rcpm}
\renewcommand{\arraystretch}{1.2}
\begin{tabular}{cccccc}
PDF set   & $\Rcpm$ & $\Delta_{\text{PDF}}$ & $\Delta_{\text{scales}}$ & $\Delta_{\alpS}$ & Total uncert.\\
\hline
MSHT20   & $0.948$ & $^{+0.021}_{-0.011}$ & $\pm0.001$ & $\pm0.0001$  & $^{+0.021}_{-0.011}$ \\
CT18           & $0.955$ & $\pm 0.003$ & $\pm 0.003$ & $\pm0.001$ & $\pm 0.003$ \\
CT18Z          & $0.958$ & $\pm 0.003$ & $\pm0.001$ & $\pm0.001$ & $\pm0.003$ \\ 
ABMP16    & $0.964$ & $\pm 0.002$ & $\pm0.001$ & $\pm0.001$ & $\pm0.002$ \\ 
NNPDF3.0           & $0.935$ & $\pm 0.017$ & $\pm0.001$ & $\pm0.0001$ & $\pm0.017$ \\
NNPDF3.1           & $0.939$ & $\pm 0.020$ & $\pm0.001$ & $\pm0.0001$ & $\pm0.020$ \\[\cmsTabSkip]

\multicolumn{6}{c}{CMS: $0.950 \pm 0.005\stat \pm 0.010\syst$}  \\
\end{tabular}
\end{table*}

\begin{figure}[htbp!]
\centering
     \includegraphics[width=0.49\textwidth]{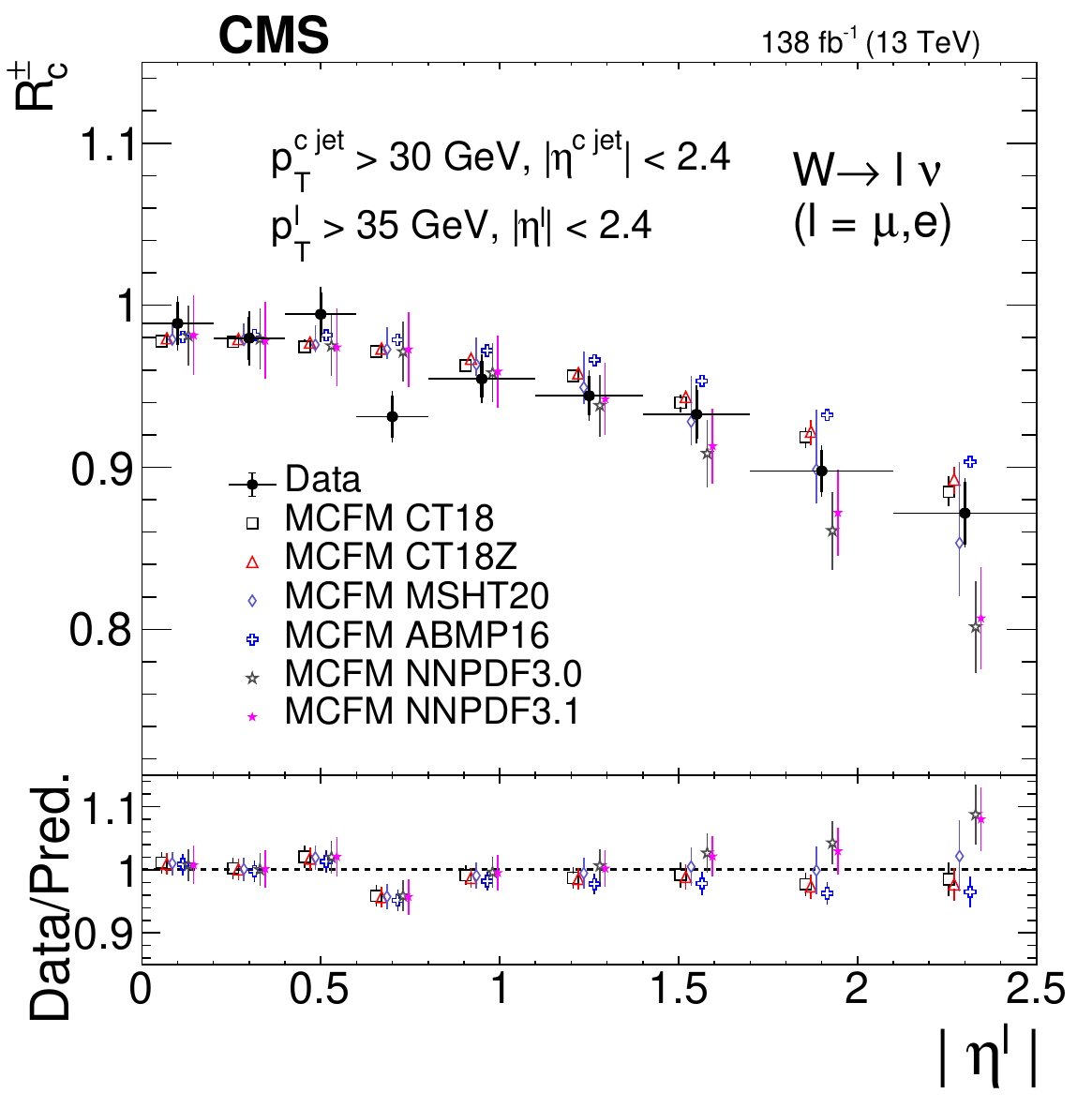}
     \includegraphics[width=0.49\textwidth]{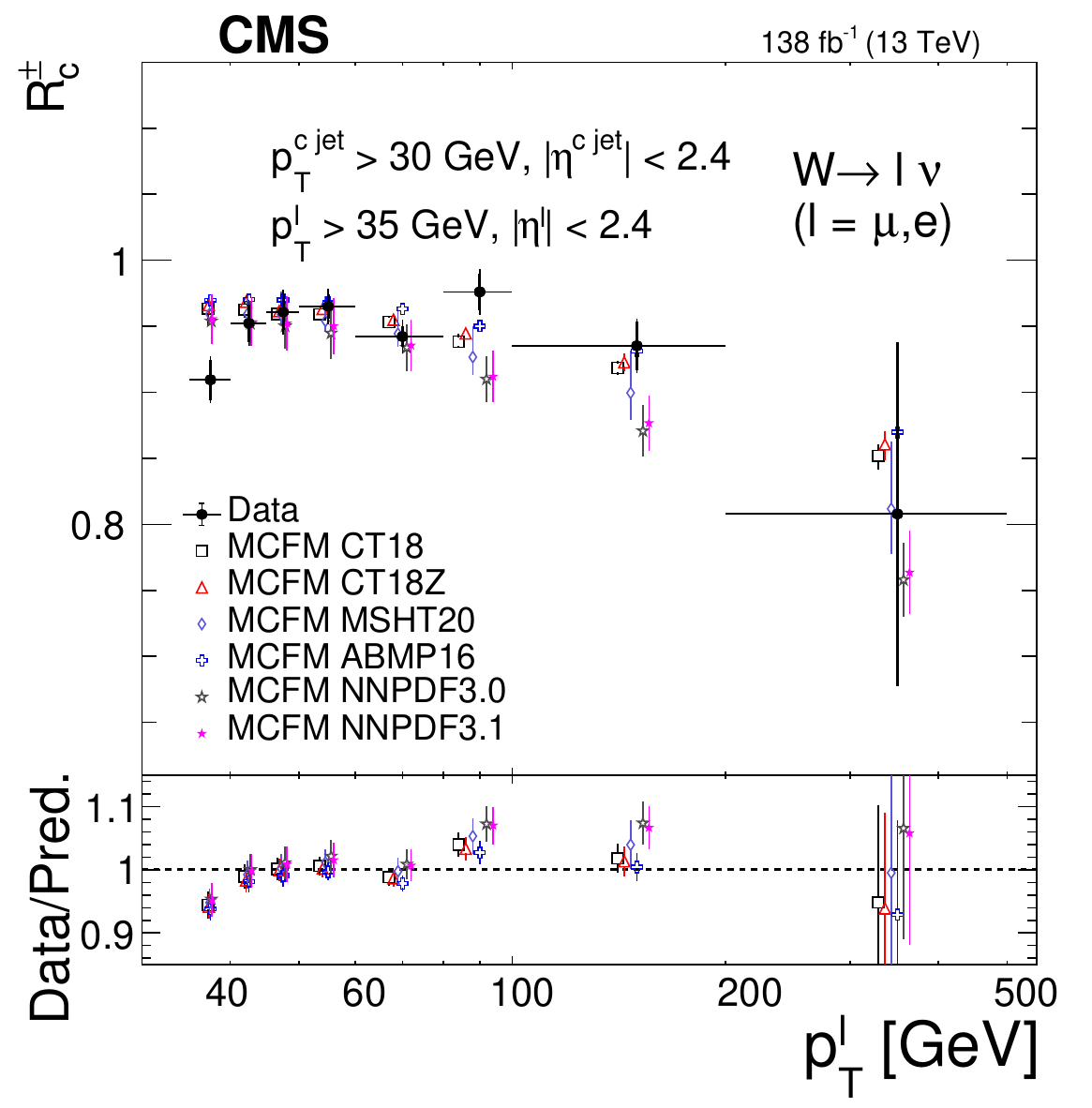}
    \caption{Measured cross section ratio $\Rcpm$ as a function of the absolute value of $\eta^\ell$ (\cmsLeft) and $\pt^\ell$ (\cmsRight), compared with
    the NLO QCD $\MCFM$ calculations using different NLO PDF sets. Error bars on data points include statistical and systematic uncertainties. Symbols showing the theoretical expectations are slightly displaced in the horizontal axis for better visibility. The ratios of data to predictions are shown in the lower panels. The uncertainty in the ratio includes the uncertainties in both data and prediction.
}
    \label{fig:Rcpm_diff}
\end{figure}

The cross section ratio $\Rcpm$ is also measured differentially as a function of $\abs{\eta^\ell}$ and $\pt^\ell$. The measurements are compared with the
$\MCFM$ predictions in Fig.~\ref{fig:Rcpm_diff}. 
The predictions are generally consistent with the measurements, with some small deviations in shape within 5\%. The cross section ratio decreases with $\abs{\eta^\ell}$ from $\Rcpm\sim1$ in the central region to about 0.87 for the most forward lepton pseudorapidity values. This behaviour is expected since different Bjorken $x$ regions are being probed. At larger $x$ values, corresponding to higher values of $\abs{\eta^\ell}$, $\PWmc$ production increases relative to $\PWpc$ because of the growing contribution initiated by the valence down quark. 
The differences between the predictions made using PDF sets with and without strange quark asymmetry grow with increasing $\abs{\eta^\ell}$ and $\pt^\ell$. However, with the current uncertainties, the data cannot distinguish between both sets of predictions. 

\section{Comparison with predictions using NNLO QCD and NLO EW calculations~\label{sec:xsec_NNLO}}
The first computation of NNLO QCD corrections for $\Wc$ production has recently been presented~\cite{NNLOWc,NNLOWc-ext}. 
The latest calculations include full off-diagonal CKM dependence up to NNLO QCD accuracy, and the dominant NLO EW corrections. In addition, a modified anti-\kt jet algorithm (flavored anti-\kt~\cite{flavour-anti-kt}) is used to guarantee that the computations are infrared safe.  This is important for a fair comparison between theory predictions and experimental measurements, since experimental results are derived using the anti-\kt jet algorithm.

{\tolerance=800
Predictions corresponding to the phase space of the CMS measurements presented in this paper, $\pt^{\ell}>35\GeV$, $\abs{\eta^{\ell}}<2.4$, $\pt^{\cjet}>30\GeV$, $\abs{\eta^{\cjet}}<2.4$, $\Delta R ({\text{jet}},\ell)>0.4$, have been specifically computed  for the purpose of this comparison, using the charge-dependent flavored anti-\kt jet algorithm with parameter $a=0.1$, and the same input parameters as in  Ref.~\cite{NNLOWc-ext}. The theoretical cross sections are provided at LO, NLO, and NNLO QCD accuracies. At LO, the  $\Wc$ process is defined at order $\mathcal{O}(\alpS\alpha^2)$ in the strong and EW couplings.  At NLO, the QCD corrections include all virtual and real contributions of order $\mathcal{O}(\alpS^2\alpha^2)$. In the same way, at NNLO accuracy all double-virtual, double-real, and real-virtual contributions of order $\mathcal{O}(\alpS^3\alpha^2)$ are included. The calculation is carried out in the 5-flavor scheme with massless bottom and charm quarks. NLO EW corrections of order $\mathcal{O}(\alpS\alpha^3)$ are calculated including all virtual corrections and the real corrections involving single real photon emission to cancel the corresponding IR divergences appearing in the EW one-loop amplitude. 
\par}

{\tolerance=800 
The nominal renormalization and factorization scales are set both to $\frac{1}{2}(E_\text{T,\PW}+\pt^{\cjet})$, where $E_\text{T,\PW}=\sqrt{\smash[b]{M_{\PW}^2+(\vec{p}_{\text{T}}^{\ell}+\vec{p}_{\text{T}}^{ \PGn})^2}}$.  To estimate missing higher-order QCD corrections, the scale uncertainty is obtained by independently varying the two scales by factors of 0.5, 1, 2, and taking the envelope of the predictions obtained with all variations excluding the cases where one scale is reduced and the other is increased at the same time. 
\par}

The calculation was performed for the most representative PDF set, which allows for strange asymmetry, NNPDF3.1. The NNLO QCD PDF set was used for computing the predictions for all orders,  following the PDF4LHC recommendation~\cite{PDF4LHC}. To evaluate the PDF uncertainty of the NNPDF3.1 sets, specialized minimal PDF sets~\cite{PDF-minimal}, which contain only 8 replicas, were used. The PDF uncertainty is calculated as the square root of the quadratic sum of the differences between the cross section obtained with the nominal PDF and that obtained with each replica. 

In Table~\ref{table:NNLOincl},  the theoretical predictions for the OS, SS, and \OSSS inclusive fiducial cross section are given at LO, NLO, and NNLO QCD accuracies.  The QCD corrections show good perturbative convergence, since the NNLO QCD corrections are significantly smaller than the NLO ones. The NNLO correction for the \OSSS cross section is negative, about -2\%. This occurs becausse the NNLO QCD corrections to SS are larger than those for OS; at LO there is no SS contribution to the $\Wc$ process and the first SS contribution enters at NLO.
The cross section calculated at NNLO QCD including NLO EW corrections is also shown in Table~\ref{table:NNLOincl}. The EW corrections amount to -2\%. They were included as a multiplicative factor with negligible statistical uncertainty.  

\begin{table*}[htbp]
\centering
 \topcaption{
Predictions for $\SWc$ in the phase space of the analysis.
For each QCD and EW order, the central values of the OS, SS and \OSSS predictions are given, together with the statistical, scales, PDF, and total uncertainties of the \OSSS prediction. All values are given in pb. The last row in the table gives the experimental result presented in this paper.}
\label{table:NNLOincl}
\begin{tabular}{ccccccccc}
QCD order & EW order &  $\sigma_{\Wc}^{\text{OS}}$ & $\sigma_{\Wc}^{\text{SS}}$  & $\sigma_{\Wc}^{\OSSS}$ & $\Delta_{\text{stat}}^{\OSSS}$ & $\Delta_{\text{scales}}^\OSSS$ & $\Delta_{\text{PDF}}^{\OSSS}$ & $\Delta_{\text{Total}}^{\text{\OSSS}}$ \\ 
\hline
LO  & LO & 137.4  & 0  & 137.4  & $\pm 0.1$  & $^{+16.6}_{-13.3}$  & $\pm 5.1$  & $^{+17.4}_{-14.3}$ \\
NLO  & LO & 182.4  & 4.1  & 178.3  & $\pm 0.3$ & $^{+9.3}_{-9.4}$  & $\pm 6.8$  &  $^{+11.6}_{-11.6}$\\
NNLO  & LO & 182.9  & 8.2  & 174.7  & $\pm 1.0$ & $^{+1.2}_{-2.8}$  & $\pm 6.8$  & $^{+7.0}_{-7.4}$ \\
NNLO  & NLO & 179.1  & 8.0  & 171.1  & $\pm 1.0$ & $^{+1.2}_{-2.8}$  & $\pm 6.8$  & $^{+7.0}_{-7.4}$ \\[\cmsTabSkip]

\multicolumn{9}{c}{CMS: $163.4 \pm 0.5\stat \pm 6.2\syst\unit{pb}$}  \\ 
\end{tabular}
\end{table*}

At LO and NLO the total uncertainty in the predictions is dominated by the scale uncertainty (around 5\% at NLO). At NNLO the scale uncertainty is reduced to 1\%, and the PDF uncertainty (4\%) dominates.  The inclusion of NNLO QCD corrections provides a more precise  determination of the strange quark  content of the proton from the cross section observable.   

\begin{figure}[htbp!]
\centering
\includegraphics[width=0.5\textwidth]{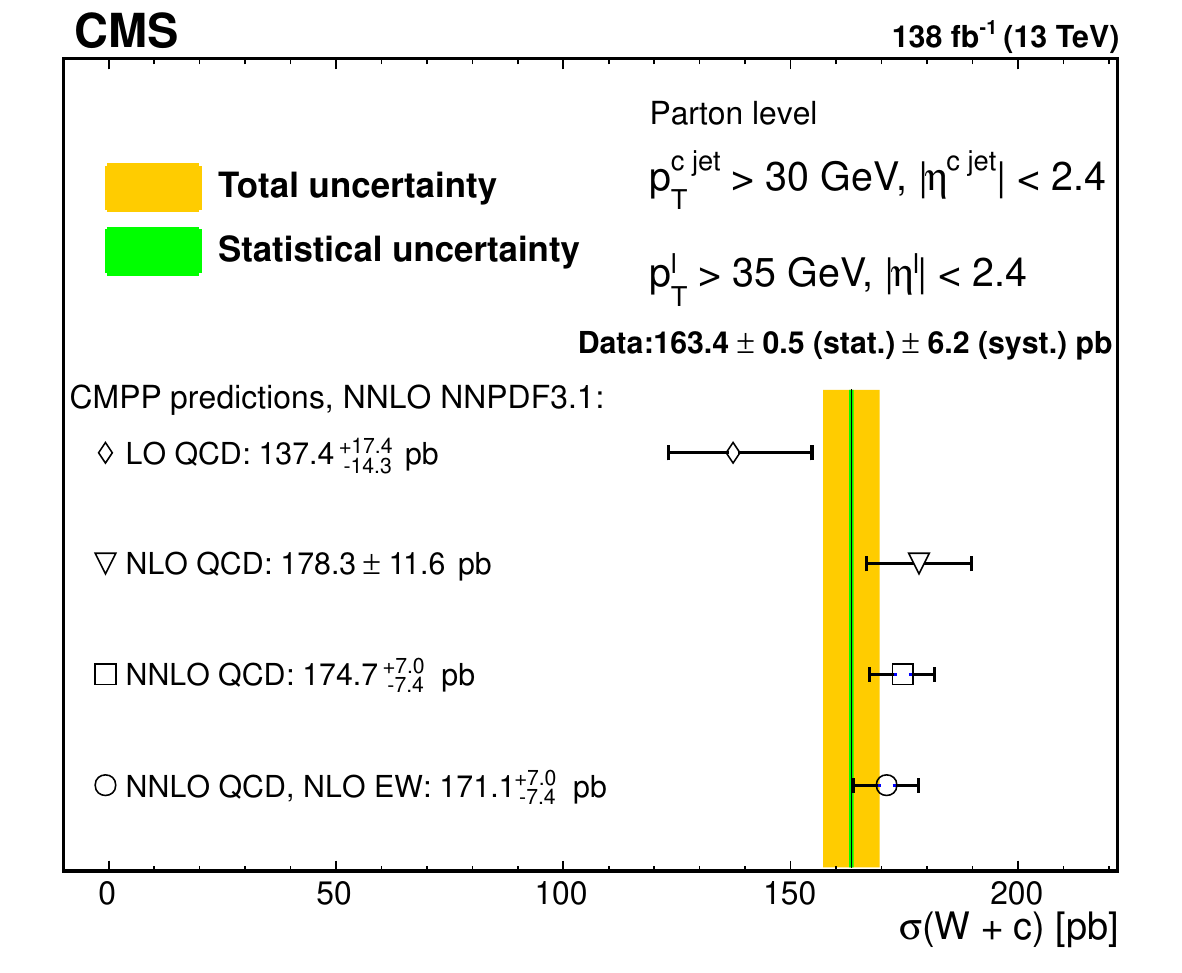}
\caption{
Comparison of the experimental measurement of $\SWc$ with the \OSSS LO, NLO, and NNLO QCD predictions, and NLO EW corrections. The NNLO QCD NNPDF3.1 PDF set is used for computing all the predictions. CMPP stands for the authors of the calculations~\cite{NNLOWc-ext}. Horizontal error bars indicate the total uncertainty in the predictions.
}
\label{fig:xs-NNLO}
\end{figure}

The \OSSS predictions are compared with the fiducial cross section measurement in Fig.~\ref{fig:xs-NNLO}. 
The \OSSS subtraction reduces the NNLO corrections, but does not remove them completely. The inclusion of the NNLO corrections decreases the uncertainty in the prediction and also brings it closer to the experimental measurement. The EW NLO corrections further improves the agreement between the theoretical prediction and experimental data. The theoretical prediction and the experimental measurement agree within uncertainties.   

No efficiency correction has been applied to account for the different flavor assignments in the jet algorithms of the predictions (flavored anti-\kt) and the experimental measurements (anti-\kt). In Ref.~\cite{NNLOWc-ext} the difference in the predictions from the standard anti-\kt and the flavored anti-\kt algorithms is studied. Due to the lack of flavored infrared safety for the standard anti-\kt algorithm, such a comparison can be done only at NLO with the help of a parton shower. The difference in the fiducial cross section predictions is below 1\%. Similarly, the effect in the NNLO theoretical $\Wc$ cross section prediction using variations of the flavored anti-\kt algorithm, and the flavored \kt algorithm is studied. Differences are also below 1\%.

\begin{figure}[htbp!]
\centering
     \includegraphics[width=0.49\textwidth]{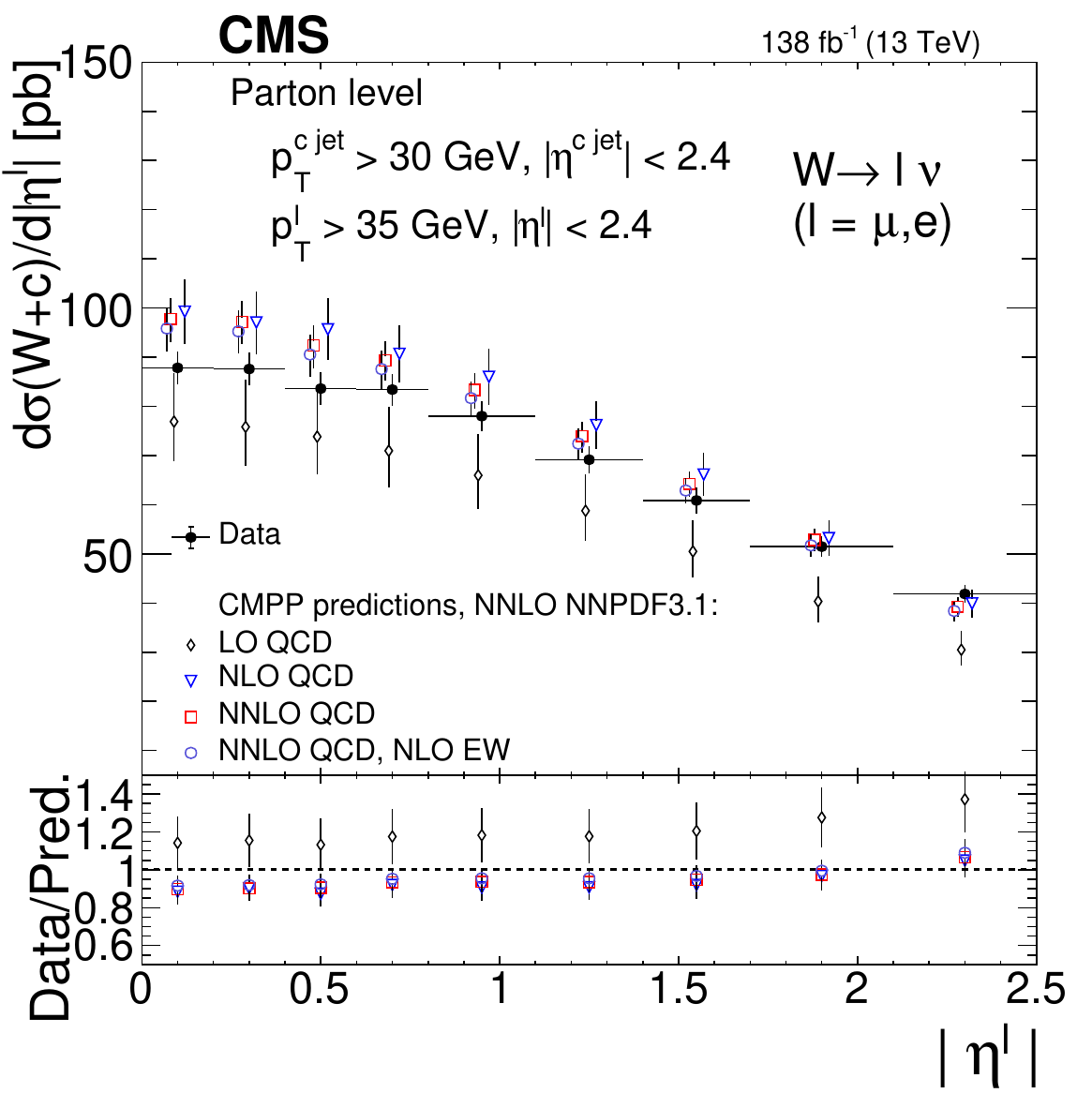}
     \includegraphics[width=0.49\textwidth]{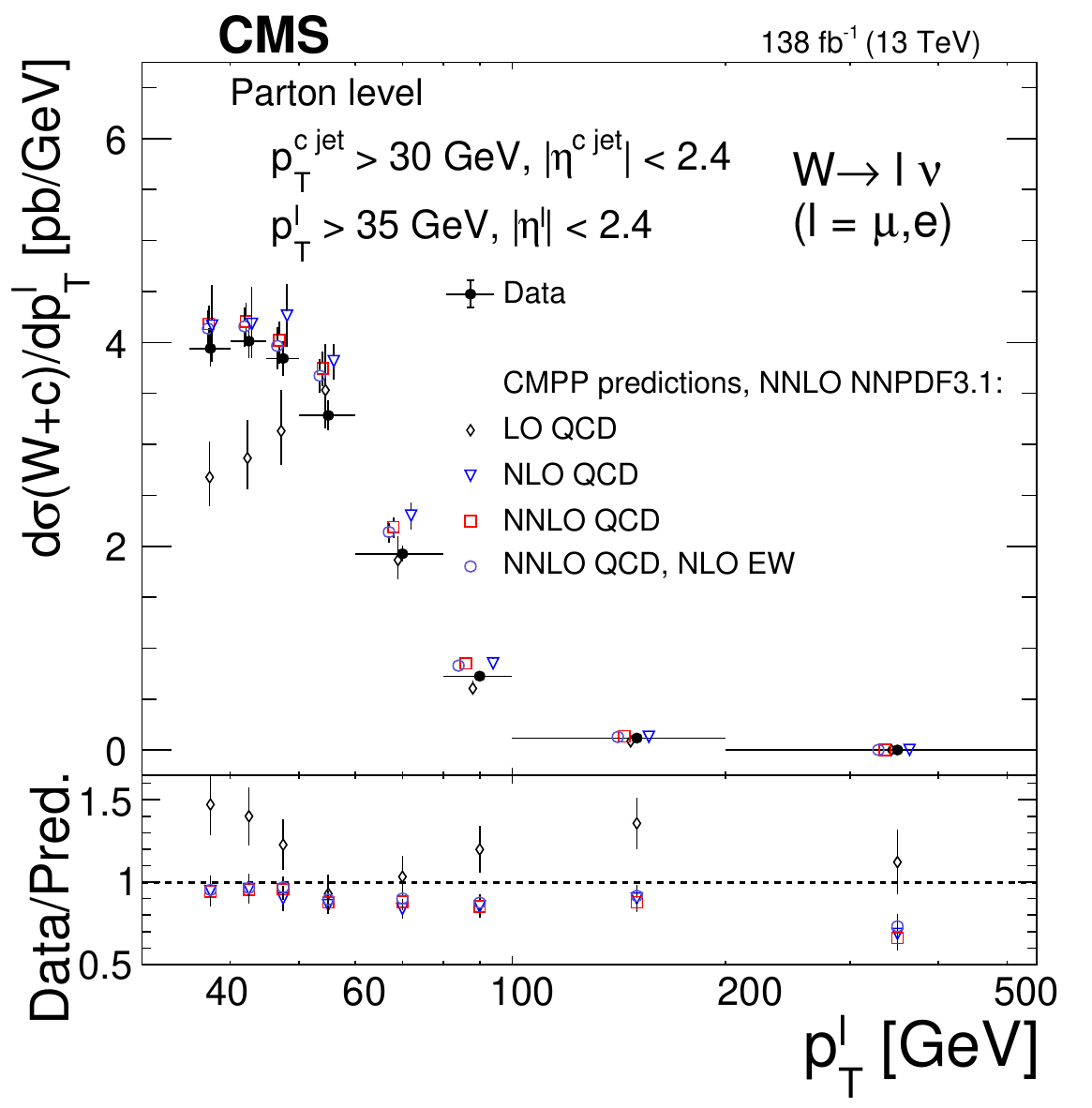}
    \caption{Comparison of the measured differential cross sections $\SWcdifflineeta$ (\cmsLeft) and $\SWcdifflinept$ (\cmsRight) with the \OSSS LO, NLO, and NNLO QCD predictions, and NLO EW corrections. The NNLO QCD NNPDF3.1 PDF set is used for computing all the predictions. CMPP stands for the authors of the calculations~\cite{NNLOWc-ext}. Error bars on data points include statistical and systematic uncertainties. Symbols showing the theoretical expectations are slightly displaced in the horizontal axis for better visibility. The ratios of data to predictions are shown in the lower panels. The uncertainty in the ratio includes the uncertainties in both data and prediction. 
}
\label{fig:xs-NNLO-diff}
\end{figure}

The predictions are also compared with the differential cross section measurements $\SWcdifflineeta$ and  $\SWcdifflinept$ in Fig.~\ref{fig:xs-NNLO-diff}. The NLO correction is approximately flat in  $\abs{\eta^\ell}$ while it is larger at low and high values of $\pt^\ell$. The NLO predictions are very similar to those shown in Fig.~\ref{fig:Sc_w_th_parton} calculated with  $\MCFM$ at NLO  using the same PDF set (NNPDF3.1). The NNLO correction is small and does not change the shape of the NLO predictions. The EW NLO correction is flat in $\abs{\eta^\ell}$ and gets larger with $\pt^\ell$, from 0.99 in the first bin to 0.90 in the highest $\pt^\ell$ bin. 

Predictions for the \OSSS cross section ratio $\Rcpm$ have also been computed and are collected in Table~\ref{table:NNLO_Rcpm}. In computing the scale variation of $\Rcpm$, the scale uncertainty for the positive and negative signatures is taken as correlated. The $\Rcpm$ observable is rather stable under perturbative QCD corrections, varying by less than 1\% from LO to NNLO accuracy. The NLO EW correction does not affect $\Rcpm$, the change being smaller than 0.1\%. 

\begin{figure}[htbp!]
\centering
\includegraphics[width=0.5\textwidth]{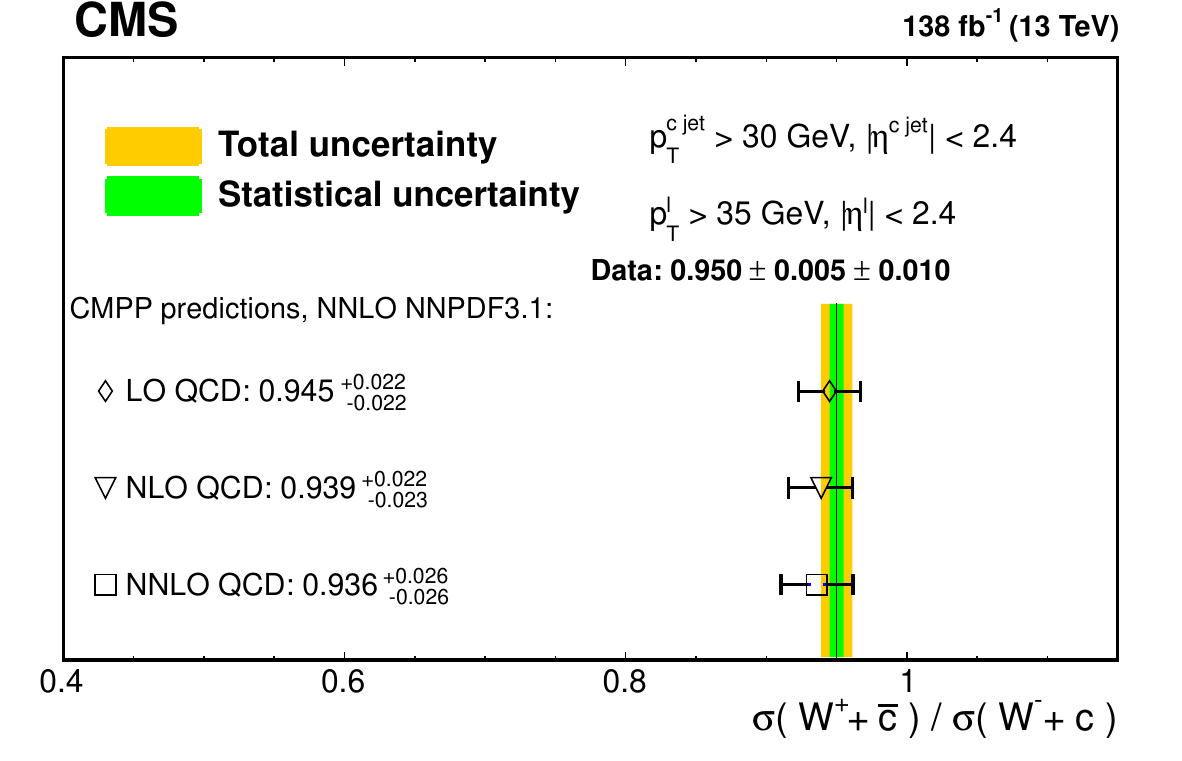}
\caption{
Comparison of the experimental measurement of $\Rcpm$ with the \OSSS LO, NLO and NNLO QCD predictions. The NNLO QCD NNPDF3.1 PDF set is used for computing all the predictions. CMPP stands for the authors of the calculations~\cite{NNLOWc-ext}. Horizontal error bars indicate the total uncertainty in the predictions.}
\label{fig:r-NNLO}
\end{figure}

\begin{figure}[htbp!]
\centering
     \includegraphics[width=0.49\textwidth]{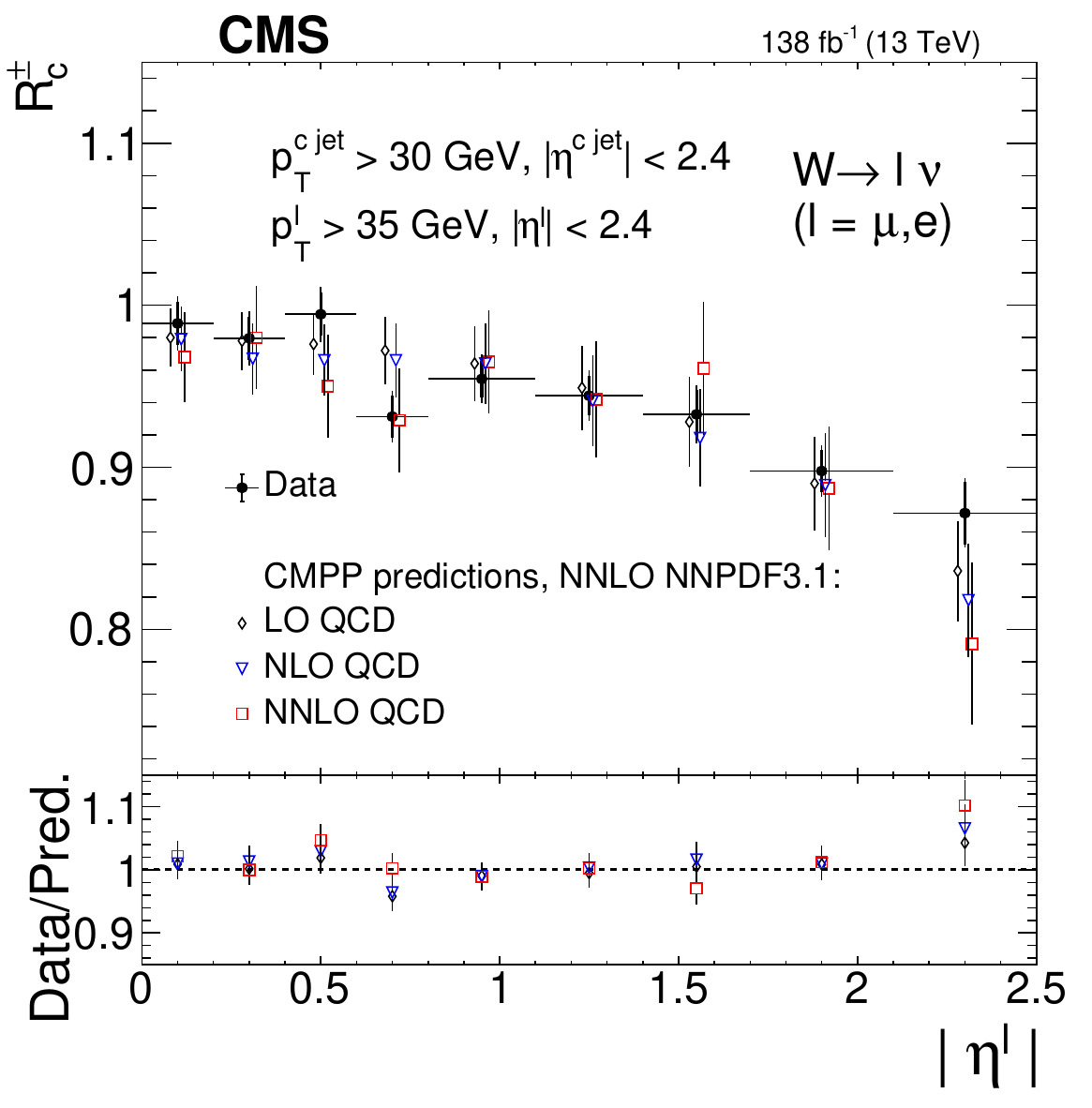}
     \includegraphics[width=0.49\textwidth]{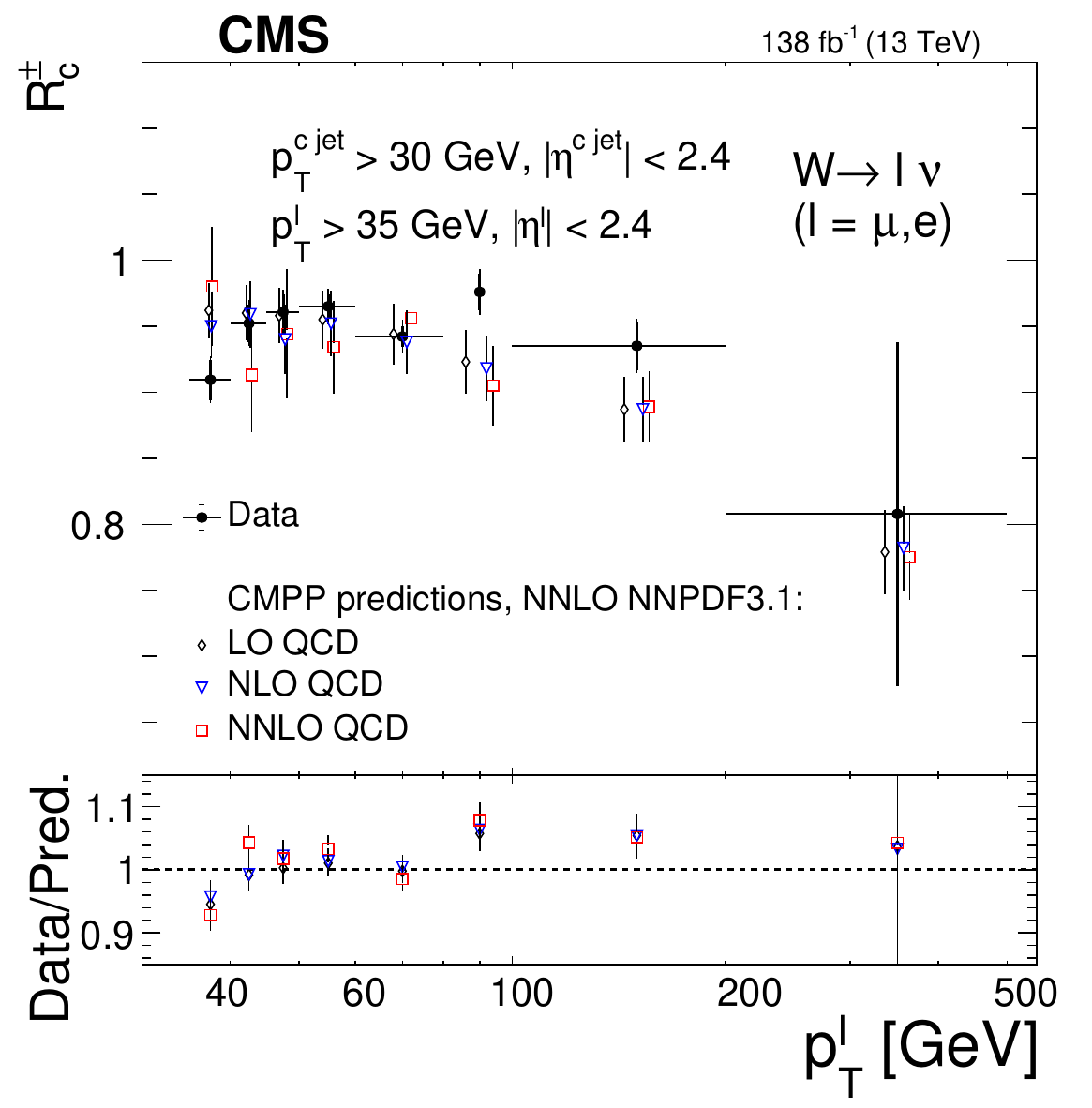}
    \caption{Comparison of the measured differential cross section ratio $\Rcpm$ as a function of the absolute value of $\eta^\ell$ (\cmsLeft) and $\pt^\ell$ (\cmsRight) with the \OSSS LO, NLO, and NNLO QCD predictions.  The NNLO QCD NNPDF3.1 PDF set is used for computing all the predictions. CMPP stands for the authors of the calculations~\cite{NNLOWc-ext}. Error bars on data points include statistical and systematic uncertainties. Symbols showing the theoretical expectations are slightly displaced in the horizontal axis for better visibility. The ratios of data to predictions are shown in the lower panels. The uncertainty in the ratio includes the uncertainties in the data and prediction.}
\label{fig:r-NNLO-diff}
\end{figure}

The comparison of the predictions with the fiducial inclusive and differential measurements are presented in Figs.~\ref{fig:r-NNLO} and~\ref{fig:r-NNLO-diff}. The inclusion of the NNLO QCD correction does not change the good agreement already observed with the predictions at NLO. 

\begin{table}[htbp]
\centering
 \topcaption{Theoretical predictions for $\Rcpm$. For each QCD order, the central values are given, together with the MC statistical, scales, PDF, and total uncertainties. The last row in the table gives the experimental result presented in this paper. 
 }
\label{table:NNLO_Rcpm}
\begin{tabular}{cccccc}
QCD order  & $\Rcpm$ & $\Delta_{\mathrm{stat}}$ & $\Delta_{\mathrm{scales}}$ & $\Delta_{\mathrm{PDF}}$  & $\Delta_{\mathrm{Total}}$ \\
\hline
LO  & 0.945 & $\pm 0.001$  & $\pm 0.001$  & $\pm 0.022$  &  $\pm 0.022$ \\
NLO  & 0.939  & $\pm 0.004$ & $\pm 0.002$   & $\pm 0.023$  &  $\pm 0.023$ \\
NNLO  & 0.936  & $\pm 0.011$ & $\pm 0.002$  & $\pm 0.023$  &  $\pm 0.026$ \\[\cmsTabSkip]

\multicolumn{6}{c}{CMS: $0.950 \pm 0.005\stat\pm 0.010\syst$}  \\
\end{tabular}

\end{table}

\section{Summary}
\label{sec:summary}

The associated production of a \PW boson with a charm quark ($\Wc$) in proton-proton ($\Pp\Pp$) collisions at a center-of-mass energy of 13\TeV was studied with a data sample collected by the CMS experiment corresponding to an integrated luminosity of $138\fbinv$.
The $\Wc$ process is selected based on the presence of a high transverse momentum lepton (electron or muon), coming from 
a \PW boson decay, and a jet with the signature of a charm hadron decay. 
Charm hadron decays are identified either by the presence of a muon inside a jet or by reconstructing a secondary decay vertex within the jet. Measurements are combined from the four different channels: electron and muon \PW boson decay channels, muon and secondary vertex charm identification channels.

Cross section measurements, within a fiducial region defined by the kinematics of the lepton from the \PW boson decay and the jet originated by the charm quark ($\pt^{\ell}>35\GeV$, $\abs{\eta^{\ell}}<2.4$, $\pt^{\cjet}>30\GeV$, $\abs{\eta^{\cjet}}<2.4$), are unfolded to the particle and parton levels. Cross sections are also measured differentially, as functions of $\abs{\eta^\ell}$ and $\pt^\ell$. The cross section ratio for the processes $\PWpc$ and $\PWmc$ is measured as well, achieving the highest precision in this measurement to date. 

The measured fiducial $\SWc$ production cross section unfolded to the particle level is:
\begin{linenomath*}
\ifthenelse{\boolean{cms@external}}{
\begin{multline*}
\sigma(\ppWc){\mathcal{B}}(\PW \to \ell \PGn)\\ 
= 148.7 \pm 0.4 \stat \pm 5.6 \syst \unit{pb}.
\end{multline*}
}{
\begin{equation*}
\sigma(\ppWc){\mathcal{B}}(\PW \to \ell \PGn) = 148.7 \pm 0.4\stat \pm 5.6\syst\unit{pb}.
\end{equation*}
}
\end{linenomath*}
The cross section measurement unfolded to the parton level yields:
\begin{linenomath*}
\ifthenelse{\boolean{cms@external}}{
\begin{multline*}
\sigma(\ppWc){\mathcal{B}}(\PW \to \ell \PGn)\\ 
= 163.4 \pm 0.5 \stat \pm 6.2 \syst \unit{pb}.
\end{multline*}
}{
\begin{equation*}
\sigma(\ppWc){\mathcal{B}}(\PW \to \ell \PGn) = 163.4 \pm 0.5\stat \pm 6.2\syst\unit{pb}.
\end{equation*}
}
\end{linenomath*}
The measured $\SWpc/\SWmc$ cross section ratio is:
\begin{linenomath*}
\begin{equation*}
\frac{\sigma(\ppWpc)}{\sigma(\ppWmc)} = 0.950 \pm 0.005\stat \pm 0.010 \syst.
\end{equation*}
\end{linenomath*}

{\tolerance 9000
The measurements are compared with theoretical predictions. The particle level measurements are compared with the predictions of the \MGvATNLO MC generator.  The parton level cross section measurements are compared with NLO QCD calculations from the $\MCFM$ program using different PDF sets and with recently available NNLO QCD calculations including NLO EW corrections. 
The predicted fiducial cross section and cross section ratio are consistent with the measurements within uncertainties. The NNLO QCD and NLO EW corrections improve the agreement between the predicted and measured cross sections. Despite the improvement in precision of the cross section ratio measurement compared with previous studies, discrimination between predictions using symmetric or asymmetric strange quark and antiquark PDFs would require a further reduction of experimental and theoretical uncertainties. 
The theoretical uncertainty is dominated by the PDF uncertainties. The inclusion of the cross section measurements in future PDF fits should improve the modeling of the  strange parton distribution function of the proton.
\par}

\begin{acknowledgments}
We congratulate our colleagues in the CERN accelerator departments for the excellent performance of the LHC and thank the technical and administrative staffs at CERN and at other CMS institutes for their contributions to the success of the CMS effort. In addition, we gratefully acknowledge the computing centers and personnel of the Worldwide LHC Computing Grid and other centers for delivering so effectively the computing infrastructure essential to our analyses. Finally, we acknowledge the enduring support for the construction and operation of the LHC, the CMS detector, and the supporting computing infrastructure provided by the following funding agencies: SC (Armenia), BMBWF and FWF (Austria); FNRS and FWO (Belgium); CNPq, CAPES, FAPERJ, FAPERGS, and FAPESP (Brazil); MES and BNSF (Bulgaria); CERN; CAS, MoST, and NSFC (China); MINCIENCIAS (Colombia); MSES and CSF (Croatia); RIF (Cyprus); SENESCYT (Ecuador); MoER, ERC PUT and ERDF (Estonia); Academy of Finland, MEC, and HIP (Finland); CEA and CNRS/IN2P3 (France); BMBF, DFG, and HGF (Germany); GSRI (Greece); NKFIH (Hungary); DAE and DST (India); IPM (Iran); SFI (Ireland); INFN (Italy); MSIP and NRF (Republic of Korea); MES (Latvia); LAS (Lithuania); MOE and UM (Malaysia); BUAP, CINVESTAV, CONACYT, LNS, SEP, and UASLP-FAI (Mexico); MOS (Montenegro); MBIE (New Zealand); PAEC (Pakistan); MES and NSC (Poland); FCT (Portugal); MESTD (Serbia); MCIN/AEI and PCTI (Spain); MOSTR (Sri Lanka); Swiss Funding Agencies (Switzerland); MST (Taipei); MHESI and NSTDA (Thailand); TUBITAK and TENMAK (Turkey); NASU (Ukraine); STFC (United Kingdom); DOE and NSF (USA).
\hyphenation{Rachada-pisek} Individuals have received support from the Marie-Curie program and the European Research Council and Horizon 2020 Grant, contract Nos.\ 675440, 683211, 724704, 752730, 758316, 765710, 824093, 884104, and COST Action CA16108 (European Union); the Leventis Foundation; the Alfred P.\ Sloan Foundation; the Alexander von Humboldt Foundation; the Science Committee, project no. 22rl-037 (Armenia); the Belgian Federal Science Policy Office; the Fonds pour la Formation \`a la Recherche dans l'Industrie et dans l'Agriculture (FRIA-Belgium); the Agentschap voor Innovatie door Wetenschap en Technologie (IWT-Belgium); the F.R.S.-FNRS and FWO (Belgium) under the ``Excellence of Science -- EOS" -- be.h project n.\ 30820817; the Beijing Municipal Science \& Technology Commission, No. Z191100007219010; the Ministry of Education, Youth and Sports (MEYS) of the Czech Republic; the Shota Rustaveli National Science Foundation, grant FR-22-985 (Georgia); the Deutsche Forschungsgemeinschaft (DFG), under Germany's Excellence Strategy -- EXC 2121 ``Quantum Universe" -- 390833306, and under project number 400140256 - GRK2497, RTG2044, INST 39/963-1 FUGG (bwForCluster NEMO), and 396021762 - TRR 257, and the state of Baden-W\"urttemberg through bwHPC (Germany); the Hellenic Foundation for Research and Innovation (HFRI), Project Number 2288 (Greece); the Hungarian Academy of Sciences, the New National Excellence Program - \'UNKP, the NKFIH research grants K 124845, K 124850, K 128713, K 128786, K 129058, K 131991, K 133046, K 138136, K 143460, K 143477, 2020-2.2.1-ED-2021-00181, and TKP2021-NKTA-64 (Hungary); the Council of Science and Industrial Research, India; the Latvian Council of Science; the Ministry of Education and Science, project no. 2022/WK/14, and the National Science Center, contracts Opus 2021/41/B/ST2/01369 and 2021/43/B/ST2/01552 (Poland); the Funda\c{c}\~ao para a Ci\^encia e a Tecnologia, grant CEECIND/01334/2018 (Portugal); the National Priorities Research Program by Qatar National Research Fund; MCIN/AEI/10.13039/501100011033, ERDF ``a way of making Europe", and the Programa Estatal de Fomento de la Investigaci{\'o}n Cient{\'i}fica y T{\'e}cnica de Excelencia Mar\'{\i}a de Maeztu, grant MDM-2017-0765 and Programa Severo Ochoa del Principado de Asturias (Spain); the Chulalongkorn Academic into Its 2nd Century Project Advancement Project, and the National Science, Research and Innovation Fund via the Program Management Unit for Human Resources \& Institutional Development, Research and Innovation, grant B05F650021 (Thailand); Isaac Newton Trust, Leverhulme Trust, and STFC grants ST/L002760/1, ST/K004883/1, and ST/T000694/1 (United Kingdom); the Kavli Foundation; the Nvidia Corporation; the SuperMicro Corporation; the Welch Foundation, contract C-1845; and the Weston Havens Foundation (USA).
\end{acknowledgments}
\bibliography{auto_generated}
\cleardoublepage \appendix\section{The CMS Collaboration \label{app:collab}}\begin{sloppypar}\hyphenpenalty=5000\widowpenalty=500\clubpenalty=5000
\cmsinstitute{Yerevan Physics Institute, Yerevan, Armenia}
{\tolerance=6000
A.~Tumasyan\cmsAuthorMark{1}\cmsorcid{0009-0000-0684-6742}
\par}
\cmsinstitute{Institut f\"{u}r Hochenergiephysik, Vienna, Austria}
{\tolerance=6000
W.~Adam\cmsorcid{0000-0001-9099-4341}, J.W.~Andrejkovic, T.~Bergauer\cmsorcid{0000-0002-5786-0293}, S.~Chatterjee\cmsorcid{0000-0003-2660-0349}, K.~Damanakis\cmsorcid{0000-0001-5389-2872}, M.~Dragicevic\cmsorcid{0000-0003-1967-6783}, A.~Escalante~Del~Valle\cmsorcid{0000-0002-9702-6359}, P.S.~Hussain\cmsorcid{0000-0002-4825-5278}, M.~Jeitler\cmsAuthorMark{2}\cmsorcid{0000-0002-5141-9560}, N.~Krammer\cmsorcid{0000-0002-0548-0985}, L.~Lechner\cmsorcid{0000-0002-3065-1141}, D.~Liko\cmsorcid{0000-0002-3380-473X}, I.~Mikulec\cmsorcid{0000-0003-0385-2746}, P.~Paulitsch, F.M.~Pitters, J.~Schieck\cmsAuthorMark{2}\cmsorcid{0000-0002-1058-8093}, R.~Sch\"{o}fbeck\cmsorcid{0000-0002-2332-8784}, D.~Schwarz\cmsorcid{0000-0002-3821-7331}, M.~Sonawane\cmsorcid{0000-0003-0510-7010}, S.~Templ\cmsorcid{0000-0003-3137-5692}, W.~Waltenberger\cmsorcid{0000-0002-6215-7228}, C.-E.~Wulz\cmsAuthorMark{2}\cmsorcid{0000-0001-9226-5812}
% ...
\par}
\vskip\cmsinstskip
\dag:~Deceased\\
$^{1}$Also at Yerevan State University, Yerevan, Armenia\\
$^{2}$Also at TU Wien, Vienna, Austria\\
% ...
\end{sloppypar}
\end{document}